**Title**

Kinetic Monte Carlo Modelling of Nano-oxide Precipitation and its Associated Stability under Neutron Irradiation for the Fe-Ti-Y-O system.

**Authors**

Chris Nellis[1], Céline Hin[1*]

[1] Department of Material Science and Engineering, Virginia Tech, Goodwin Hall, 635 Prices Fork Road – MC 0238, Blacksburg, VA 24061, USA

Corresponding Author information: Celine Hin (celhin@vt.edu)

Abstract:

While developing nuclear materials, predicting their behavior under long-term irradiation regimes spanning decades poses a significant challenge. We developed a novel Kinetic Monte Carlo (KMC) model to explore the precipitation behavior of Y-Ti-O oxides along grain boundaries within nanostructured ferritic alloys (NFA). This model also assessed the response of the oxides to neutron irradiation, even up simulated radiation damage levels in the desired long dpa range for reactor components. Our simulations investigated how temperature and grain boundary sinks influenced the oxide characteristics of a 12YWT-like alloy during heat treatments at 1023 K, 1123 K, and 1223 K. The oxide characteristics observed in our simulations were in good agreement with existing literature. Furthermore, the impact of grain boundaries on precipitation was found to be minimal. The resulting oxide configurations and positions were used in subsequent simulations that exposed them to simulated neutron irradiation to a total accumulated dose of 8 dpa at three temperatures: 673 K, 773 K, and 873 K, and at dose rates of $10^{-3}$, $10^{-4}$, and $10^{-5}$ dpa/s. This demonstrated the expected inverse relationship between oxide size and dose rate. In a long-term irradiation simulation at 873 K and $10^{-3}$ dpa/s was taken out to 66 dpa and found the oxides in the vicinity of the grain boundary were more susceptible to dissolution. Additionally, we conducted irradiation simulations of a 14YWT-like alloy to reproduce findings from neutron irradiation experiments. The larger oxides in the 14YWT-like alloy did not dissolve and displayed stability similar to the experimental results.

# 1. Introduction

The next generation of nuclear fission reactors and first-generation fusion reactors are currently under development to compete with fossil fuel sources and reduce greenhouse gas emissions. To enhance their economic viability, these new reactors are being designed with improved fuel efficiency. However, this introduces a new material challenge as these reactors need to operate at higher temperatures and in more intense irradiation environments, which surpass the capabilities of current reactor materials [1]. There is a concern that defects, such as point defects, dislocation loops, and helium bubbles resulting from irradiation damage, may weaken the materials and shorten the lifespan of these components [2]. To address these concerns, researchers have conducted extensive studies on nanostructured ferritic alloys (NFAs) for potential use as nuclear fuel cladding in reactor environments. These alloys also hold promise for applications in yet-to-be-constructed fusion reactors as materials for blanket structures [3].

NFAs are characterized by a high number density of nano-sized ($\leq$ 2 nm) particles embedded into a ferritic matrix. This study focuses on NFAs that primarily contain nano-oxide particles as opposed to carbides or nitrides. These nano-oxides provide alternative nucleation sites for the bubbles formed by transmutation He atoms produced during neutron irradiation through (n,$\alpha$) reactions. This prevents the formation of large bubbles at the metal grain boundaries, which could significantly weaken the material. Instead, smaller bubbles are formed inside the metal grain, resulting in less degradation of the material's properties [4].

Two NFA alloys, 12YWT and 14YWT, have been extensively studied and simulated, including the processing conditions required to achieve oxides of a certain size and number density [5-7]. Oxides in the NFAs typically consist of oxygen and the metal solutes Y and Ti. The exact structure of these oxides is undetermined, but they are commonly thought to have the structure $Y_2O_3$, $Y_2TiO_5$, or $Y_2Ti_2O_7$ [4]. The production of NFAs begins with a metal powder mix of all the components, with oxygen being introduced via a $Y_2O_3$ compound. The powder mix undergoes mechanical alloying to release oxygen into the solid solution. The solid solution then undergoes heat treatment through either hot isostatic pressing, hot extrusion, or spark plasma sintering [8], resulting in the formation of nano-oxides in the material [9]. Once formed, the oxides exhibit strong thermal stability at high temperatures and display little coarsening even at temperatures as high as 1273 K [10]. This thermal stability is especially relevant when the materials are intended for service temperatures below 973 K.

Another compelling characteristic of these nano-oxides is their resilience against dissolution even under intense irradiation, guaranteeing that the material retains its properties throughout its entire lifespan. However, a literature review by Wharry on experimental studies that tested the survivability of nano-oxides under intense irradiation found mixed and sometimes contradictory results across the studies [11]. Some studies found that the oxides were stable when irradiated above 873 K, while other studies showed that the oxides became less stable at that same temperature. These discrepancies are attributed to differences in material processing, the testing environment, and characterization techniques; thus, more study is required.

There are limited facilities that can produce the high-energy neutrons needed to irradiate material samples for testing. As a result, proton, electron, and ion irradiations are often used as a substitute for neutron irradiation, though the viability of this approach is a matter of debate. Ribis found that ion irradiation has a deep impact on the nano-oxide growth during irradiation, with radiation-induced oxide growth occurring 17-40 times faster than the growth in samples that were neutron irradiated [12]. Similarly, Swenson found that the stability of the nano-oxides differed when irradiated with protons and neutrons [13], with the nano-oxides being more stable under proton irradiation than neutron irradiation. These experiments also provided limited insight into the evolution of nano-oxide characteristics over the total accumulated dose, as most of the irradiation experiments only analyzed the oxides after two or three dosages.

One possible solution to address these challenges is to employ computer simulations for modeling the irradiation response [14]. The response of materials to irradiation can be simulated using computer models such as mean-field rate theory [15], molecular dynamics [16], cluster dynamics [17], and phase field models. Kinetic Monte Carlo (KMC) models have been previously used to simulate the precipitation kinetics of alloys, such as Fe-Cu [18] and Fe-Nb-C [19]. There are also models that simulate the precipitation of Y-O [20], Ti-O [21], and Y-Ti-O [22] oxides using KMC under thermal aging. Soisson extended his KMC model to allow for the modeling of electron radiation damage by introducing point defects via Frenkel Pair insertion [23].

This study develops a KMC model to simulate the NFA systems (12YWT and 14YWT) under neutron irradiation. The work builds upon a previous KMC model that simulates the precipitation of Y-Ti-O oxides in the bulk region of an NFA [22]. First, a grain boundary was incorporated into the model to provide a sink for point defects. Next, a method for simulating neutron irradiation was developed for the model. Becquart and Vincent implemented neutron irradiation in their KMC models by inserting cascade debris from molecular dynamics simulations of displacement cascades into the KMC simulation box [24,25]. Our model simulates the displacement cascade from neutron impact within the simulation box to the end of the cascade. Our code is improved over other KMC models, allowing it to mimic the displacement cascade and the resulting radiation damage up to relevant dpa levels.

This study constructs a KMC computer model that can simulate both the precipitation kinetics of the oxides in an Fe-Ti-Y-O system and the response of these oxides to neutron irradiation conditions inside a reactor. The KMC model was able to simulate neutron irradiation and the resulting cascade of displacements, reaching the relevant dpa levels seen in irradiation experiments. The study investigates the stability of oxides under varying conditions, such as temperatures and dose rates. Additionally, this study aims to compare findings regarding the thermal precipitation behavior of $Y_2Ti_2O_7$ in the bulk and near a grain boundary, the shape of the oxides produced, and the irradiation response of the system to other irradiation experiments found in the literature.

## 2. Methods

### 2.1. The Material System

The material system is represented in the KMC model similarly to the model constructed by Hin [20,21] and in a prior study of Fe-Y-Ti-O precipitation [22]. This model starts by creating a system of fixed lattice points in a bcc structure for an Fe system. These points are populated by substitutional atoms of the matrix Fe, as well as the substitutional solutes Y and Ti, while the O atoms exist on a separate octahedral sub-lattice. As noted in the prior study [22], the merged bcc unit cell with its octahedral sublattice resembles a 2x2x2 construction of simple cubic unit cells. Thus, we describe the system from the simple cubic point-of-view when discussing the thermodynamic parameters later in Section 2.2 and 2.3. The unstressed structure of the oxides differs from the bcc Fe matrix and cannot be represented in this set-up. However, considering the expected small size of these oxides, it can be assumed that the small oxides are coherent with the bcc matrix for the purposes of this study.

The vacancy and interstitial dumbbell point defects required for atomic migration of substitutional Fe, Y, and Ti atoms, reside on the substitutional lattice: the vacancy as a vacant site and the dumbbell as two atoms occupying the same lattice site. For thermodynamic considerations, all interstitial dumbbells are assumed to be oriented in the <110> position since this orientation is the most energetically favored position for the Fe-Fe dumbbell [26].

## 2.2. KMC Model

The events that the KMC model simulates include point defect generation (whether radiation-induced or thermally induced), vacancy migration, interstitial migration, and interstitialcy migration. The frequencies of these events are quantified and assigned an event frequency denoted as $\Gamma_x$, which represents the number of times per second the event is expected to occur. The model tabulates all possible events and calculates the total frequency:

$$\Gamma_{Tot} = \sum \Gamma_x \quad (1)$$

At each Monte Carlo step, a random number r[0,1] is sampled from a uniform distribution and the algorithm selects the event that corresponds to condition $\sum_0^i \Gamma_x < r\Gamma_{Tot} < \sum_0^{i+1} \Gamma_x$ . The amount of time that passes in that particular step $t_{mcs}$ is the inverse of the total event frequency:

$$t_{mcs} = \frac{1}{\Gamma_{Tot}} \quad (2)$$

The migration of interstitial dumbbells, recombination of point defects, and the treatment of grain boundary point defect annihilation for the KMC model follow the methodology laid out by Soisson [23]. This KMC model has periodic boundary conditions, and mobile defects that migrate out of the simulation box emerge at the opposite side. A perfectly planar sink is placed in the center of the simulation box to represent the grain boundary. During irradiation simulations, defect sinks, like the grain boundaries, and recombination of the vacancy and interstitial dumbbells are responsible for the removal of point defects from the simulation box. When a vacancy annihilates at the defect sink, the lattice site is filled with an atom randomly chosen from a reservoir initially seeded with six Fe atoms and whichever atom was first removed to make way for the vacancy.

When an interstitial dumbbell annihilates at the grain boundary, one atom in the dumbbell is placed into the reservoir, and the other atom from the dumbbell remains at the lattice site.

A vacancy is inserted during the thermal processes at a rate so the average vacancy concentration corresponds to the thermal equilibrium concentration. The process is described in Hin's work [20]. Point defects from irradiation are introduced through a displacement cascade from a simulated neutron strike at frequency $\Gamma_{neutStrike}$ defined in Appendix A.

The jump frequencies $\Gamma_x$ for the migration event are tied to the diffusion properties of the defects and can be calculated through the following relationship:

$$\Gamma_x = v_{Fe} \times \exp(\frac{-E_{mig}}{k_b T}) \quad (3)$$

where $v_{Fe}$ is the attempt frequency and $E_{mig}$ is the migration energy. Calculation of the migration energy in the KMC is based on the local atomic environment around the migrating species:

$$E_{mig} = e^{SP} - I_3 - I_4 - I_o - I_{PD} \quad (4)$$

$$E_{mig} = e^{SP} - \sum_j \varepsilon_{ij}^3 N_j^3 - \sum_j \varepsilon_{ij}^4 N_j^4 - \sum_{n=1,2} \varepsilon_{iO}^n N_O^n - \sum_j \varepsilon_{jV}^3 N_j^3 \quad (5)$$

The saddle-point energy, $e^{SP}$, remains constant regardless of any changes in local composition. $\varepsilon_{XY}^n$, are the pair-interaction energies between solutes XY at the *n* nearest neighbor distance, ordered 1-4 in the simple cubic system. The 3rd and 4th nearest neighbors in the simple cubic system correspond to the 1st and 2nd nearest neighbors in the bcc system. Other models only use the first nearest neighbors (in the bcc system) of the solute atom and point defect when calculating the migration energy of substitutional atoms, denoted as $I_3$ and $I_{PD}$. However, to ensure the precipitates have the correct order and stoichiometry, our model includes the 2nd nearest neighbor (in the bcc system), denoted as I4. Another addition to previous models is the inclusion of interactions of atoms on different sub-lattices as the atom is migrating, denoted as $I_O$. Equation 4 is rewritten as Equation 5 to replace the placeholder terms $I_x$ with their actual terms. $N_j^n$ is the number of *n* nearest neighbors of species *j* that the migrating atom *i* has. For example, $N_Y^3$ is the number of Y atoms in the third nearest neighbor position that the migrating atom *i* has. The oxygen atoms migrate through a similar mechanism, except that the migration is visualized as the oxygen atom hopping to an adjacent empty octahedral site. Mobile clusters are not included to prevent the complexity of needing to gather significantly more thermodynamic data. We do not believe this assumption will significantly impact the results, as it is also common in rate theory models [27].

## 2.3. Parameterization

This section provides details regarding the thermodynamic and kinetic parameterization of the Fe-Ti-Y-O system.

**Table 1**: Pair-interaction energies for Fe-Ti-Y-O system (eV) as a function of nearest neighbors in the simple cubic system.

| | **1** | **2** | **3** | **4** |
|---|---|---|---|---|
| Fe-Fe | - | - | -0.611 | -0.611 |
| Fe-Y | - | - | -0.59 | -0.52 |
| Y-Y | - | - | -0.57 | --0.69 |
| Fe-Ti | - | - | -0.65 | -0.53 |
| Ti-Y | - | - | -0.71 | -0.68 |
| Ti-Ti | - | - | -0.69 | -0.70 |
| Fe-V | - | - | -0.21 | 0.0 |
| Y-V | - | - | -0.35 | 0.0 |
| Ti-V | - | - | -0.35 | 0.0 |
| Fe-I | - | - | -0.1 | 0.0 |
| Y-I | - | - | 0.25 | 0.0 |
| Ti-I | - | - | -.1 | 0.0 |
| Fe-O | 0.00 | 0.00 | - | - |
| Y-O | -0.01 | -0.11 | - | - |
| Ti-O | -0.04 | -0.04 | - | - |
| O-O | 0.1 | -0.116 | 0.1 | -0.116 |

Table 1 lists the pair-interaction energies used to describe the Fe-Ti-Y-O system, ranging from 1st to 4th nearest neighbors in the simple cubic system. The model extends the parameterization performed in a prior precipitation study [22] by adding interstitial dumbbell interactions. Pair-interaction energies between all constituent elements are determined using a procedure described in Hin's model [20], which involves a combination of experimental and calculated material property values, including defect formation energy, solubility limits, and solubility products. The pair-interactions for yttrium and titanium with a dumbbell were derived from the energies of formation of the <110> dumbbell when the solute atom was in one of its first-nearest neighbor positions [26].

Another critical aspect for the accuracy of the model is linking the system's kinetics to the diffusion properties of each solute in the Fe system. The system is configured in such a way that each solute in pure bcc metal reproduces the diffusion values obtained from the literature or experiments, as described in a previous paper simulating the same material [22]. Table 2 provides a list of the diffusion values essential to the system.

**Table 2:** Diffusion parameters for the Fe-Ti-Y-O system.

| Input Parameter | Value | Unit | Reference |
|---|---|---|---|
| lattice parameter | $2.87 \times 10^{-10}$ | m | |
| Vacancy Migration Energy, Fe | 0.66 | eV | [28] |
| Vacancy Migration Energy,Y | 0.23 | eV | [29] |
| Vacancy Migration Energy,Ti | 0.81 | eV | [30] |
| Interstitial Migration Energy, Fe | 0.35 | eV | [23] |
| Interstitial Migration Energy, Y | 0.21 | eV | calculated |
| Interstitial Migration Energy, Ti | 0.24 | eV | calculated |
| Vacancy Formation Energy | 2.24 | eV | [20] |
| Oxygen Migration Energy | 0.48 | eV | [31] |
| Vacancy pre-exponential factor, Fe | $6.0\text{x}10^{-4}$ | $m^2/s$ | [28] |
| Vacancy pre-exponential factor, Y | $3.7\text{x}10^{-7}$ | $m^2/s$ | [29] |
| Vacancy pre-exponential factor, Ti | $2.1\text{x}10^{-1}$ | $m^2/s$ | [30] |
| Interstitial pre-exponential factor, Fe | $1.0\text{x}10^{-5}$ | $m^2/s$ | |
| Interstitial pre-exponential factor, Y | $1.0\text{x}10^{-5}$ | $m^2/s$ | |
| Interstitial pre-exponential factor, Ti | $1.0\text{x}10^{-5}$ | $m^2/s$ | |
| Interstitial attempt frequency | $1.0\text{x}10^{14}$ | /s | |

### 2.4. Treatment of the Grain Boundary

At the grain boundary (simulated here as a perfectly planar sink), segregation energies are introduced, which can influence the favorability of certain solutes to segregate at the grain boundary. Each solute species has a specific segregation energy that is included in the calculation of the migration energy when the solute is on the plane of the grain boundary. There is a slight temperature dependence of the segregation energies due to entropy. These values are listed in Table 3.

**Table 3:** Segregation enthalpy and entropy of selected solutes in bcc Fe from Lejcek [32].

| | O | Ti | Y |
|---|---|---|---|
| **ΔH** | 0.9984 eV | 0.3328 eV | 0.6448 eV |
| **ΔS** | $4.784 \times 10^{-4}$ eV/K | $3.016 \times 10^{-4}$ eV/K | $6.552 \times 10^{-4}$ eV/K |

The O and Ti segregation values are based on calculations by Lejcek, while the Y segregation values were determined for this study from the solubility limit of Y in bcc Fe using Lejcek's method, assuming a general type of grain boundary [32]. In the KMC model, segregation is treated as follows: when solute atoms are located on the same plane as the grain boundary, the segregation energies are added to the migration energies calculated in Eq. 5. These energies either repel the solute or hinder it from leaving the grain boundary, ultimately leading to the enrichment or depletion of that solute's atomic species at the grain boundary.

2.5. Neutron Irradiation

The primary source of damage during irradiation is the impact of high-energy neutrons with metal atoms in the material. This collision creates a cascade of displacements, which introduces defects such as vacancies and interstitial dumbbells into the material. Neutrons emitted from nuclear reactions in the fuel typically have energies in the 0.1 MeV to 10 MeV range [33]. To simulate the neutron impact, we assume that the neutron has an energy of 1 MeV. The neutron strikes a random atom in a substitutional site in the matrix and, due to the small box size, is assumed to exit the simulation box without depositing more energy. The atom struck is referred to as the primary knock-on atom (PKA), and all subsequently displaced atoms are called secondary knock-on atoms (SKAs). These SKAs, imparted with energy $E_{SKA}$, will then move through the metal matrix and lose energy as they collide with other atoms and exchange energy. The exchange of energy in these collisions is assumed to be elastic.

$$\begin{cases} E_{SKA} \geq E_d & \text{Dislodge new atom and make new SKA} \\ E_{SKA} < E_d & \text{Form interstitial dumbbell} \end{cases}$$

If the $E_{SKA}$ of the resting atom exceeds the threshold displacement energy $E_d$, then the resting atom is energized to vacate its lattice site on its own path and become another SKA in the cascade. Binary collision approximation (BCA) models provide a good approximation of the displacement cascade until $E_{SKA}$ drops below the $E_d$, where subsequent low-energy collisions, along with other processes not modeled in BCA, dissipate the excess energy. However, unlike BCA, this "thermal spike" region in the cascade process can be simulated using molecular dynamics (MD), and it has become common practice to simulate displacement cascades using MD [16]. These KMC models sample from a collection of MD-generated displacement cascade distributions rather than creating a unique cascade for every neutron strike since MD is cost-prohibitive for the long timescale of these irradiation studies. Unlike MD, the KMC model does not continuously recalculate the atomic interactions throughout the movement of an atom from one lattice point to another and its rigid lattice does not allow for atomic relaxation. This trade-off in precision allows the KMC model to produce cascade distributions at reduced computational expense so the model can achieve timescales that can reach irradiation doses up to 70 dpa and beyond.

Ortiz [34] has shown that the BCA could be improved by allowing recombination of low-energy SIAs with a vacant lattice site within a material-dependent "capture radius." When the energy of the SKA falls below the threshold energy and collides with a sitting atom, an interstitial dumbbell forms on that lattice site. At this point, any vacancy that falls within the capture radius instantly recombines with the dumbbell. At the end of a cascade, a small number of vacancies and dumbbells that have not recombined are left in the simulation box. In the bcc Fe system, a capture radius of approximately three lattice parameters would result in a similar quantity of free defects compared to MD, although it may not necessarily produce the same spatial distribution of surviving defects. Reducing the complex processes of vacancy-SIA recombination to a capture radius is a common practice in KMC irradiation models. Soisson [23] uses a 3-lattice parameter capture radius to trigger recombination for free-migrating defects in bcc Fe in his electron irradiation model.

A method for validating the neutron irradiation mechanism by tracking the number of displacements generated is provided in the Supporting Information, along with a description of the 'dpa' unit that measures irradiation damage.

## 2.6. Simulation Procedure

First, solid solutions of 12YWT and 14YWT compositions underwent heat treatments at various temperatures to study the impact of grain boundaries on the precipitation characteristics of the nano-oxides. Subsequently, oxide distributions created in the precipitation simulations were subjected to neutron irradiation at varying dose rates and temperatures up to 8 dpa to observe the relationship between environmental conditions and nano-oxide survivability. In a separate investigation, the KMC model's results were compared to those from an experiment for the 633K at 6.5 x $10^{-7}$ dpa/s for a 14YWT system. Finally, a long-term irradiation study (> 60 dpa) for the 12YWT system was conducted. Unless otherwise noted, three simulation runs were conducted for each set of conditions, and the results were averaged.

# 3. Results

## 3.1. Precipitation Results under Thermal Aging

### 3.1.1. Homogeneous and Heterogeneous Precipitation in the Presence of a Defect Sink

We examined the influence of grain boundaries on precipitation and compared the results to those obtained in bulk materials as carried out in a previous study [22]. The interaction of defects with the grain boundary follows methods outlined in previous work [15]. The chosen alloy is similar to the commercial alloy 12YWT and 14YWT. The atomic concentrations for both model alloys of the key solutes Y, Ti, and O in the bcc Fe matrix are as follows: (0.13 at% Y, 0.4 at% Ti, 0.18 at% O) for 12YWT and (0.08 at% Y, 0.27 at% Ti, 0.3 at% O) for 14YWT. Additional elements such as Cr and W were not included. Note the atomic concentrations of the constituent solutes do not exactly align with the expected stoichiometry of the oxide nanoparticles. Research has indicated that maintaining the Ti:Y ratio above 1 and limiting excess oxygen in the metal matrix are important for the thermal stability of the oxides at high temperatures [35]. These factors also limit the degree of coarsening in the system. Samples of another alloy, 14YWT, were heat-treated in simulation alongside the 12YWT for later use in the irradiation simulations.

A simulation box, sized to 400x100x100 lattice sites, was constructed with a simulated grain boundary in the center and populated with solutes to match the composition of the solid solution. The grain boundary was represented by a perfect planar sink perpendicular to the longitudinal direction, where vacancies were annihilated once they migrated onto the boundary plane. The simulations were conducted at three different temperatures (1023 K, 1123 K, and 1223 K) for the 12YWT alloy and at 1123 K for the 14YWT alloy, with three samples for each condition. The results from the 1123 K simulations were compared to experimental data obtained by Alinger for the 12YWT [7] and Miller for the 14YWT [6].

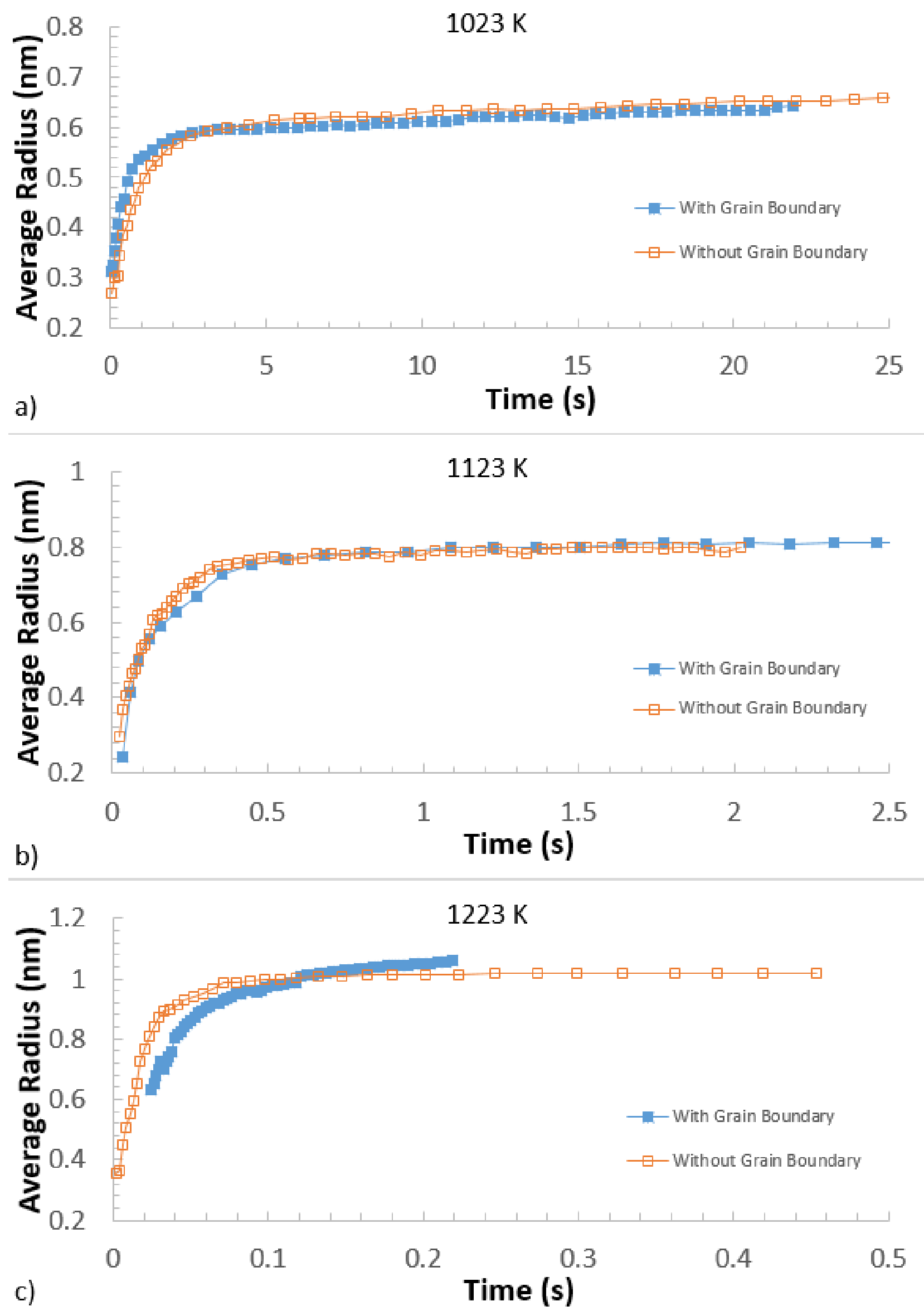

**Figure 1:** The evolution of oxide size with time with and without a grain boundary for the 12YWT composition at a) 1023K b) 1123K and c) 1223K.

Figure 1 compares the kinetic evolution at three temperatures for the 12YWT both with and without a grain boundary in the simulation box. The precipitation process begins with the rapid nucleation and growth of nano-oxides until most of the solute material is depleted from the Fe matrix, accumulating in the oxides. This is followed by a coarsening stage that is significantly slowed by the trapping of vacancies at oxide interfaces, as depicted in Fig. 2. The higher vacancy concentrations are evident at the interfaces of the precipitates, which is expected because the precipitates serve as defect sinks, capturing and retaining vacancies on their surfaces [36].

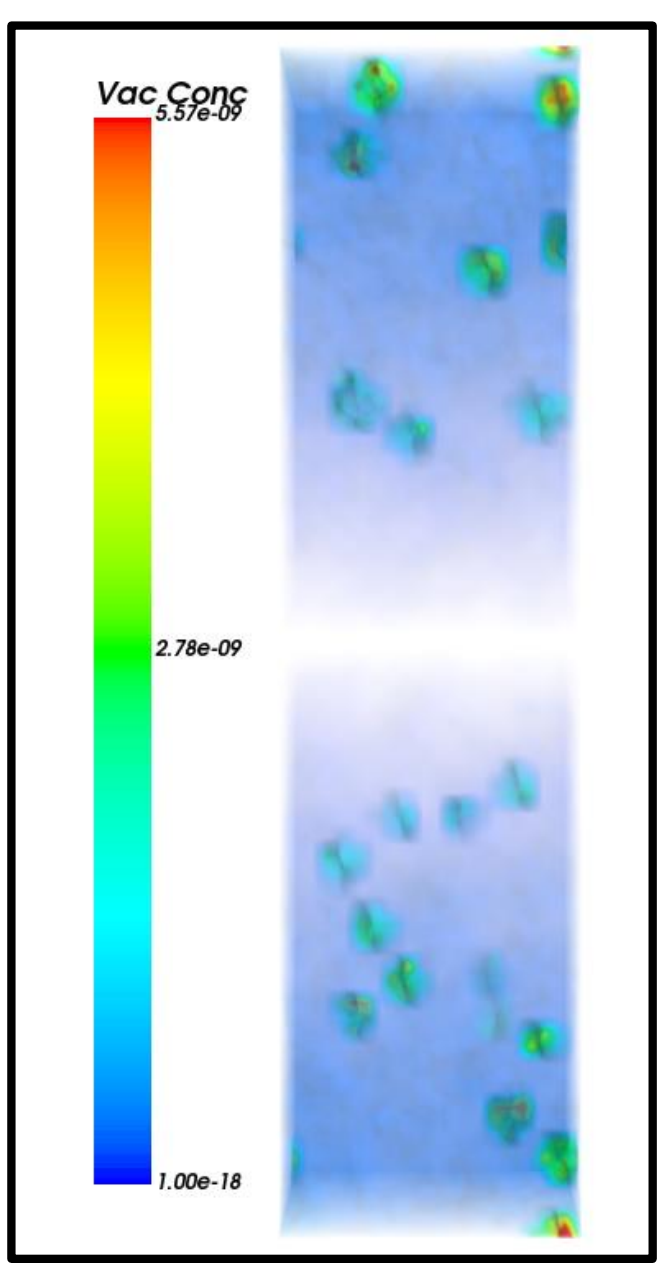

**Figure 2:** Heatmap of vacancy locations in the 12YWT-1123K with a grain boundary during heat treatment.

There was little difference in the average size of the oxides between simulations with and without a grain boundary. However, there was a difference in the oxide locations. In the bulk precipitation simulations shown in Figure 3a, the oxides were uniformly distributed. In contrast, in the grain boundary simulations, the area around the planar sink is devoid of oxide precipitates, as illustrated in Figure 3b. This precipitate-free zone extends approximately 1-2 nm away from the planar sink, and is a common phenomenon in many alloy compositions [37]. There is a noticeable decrease in vacancy concentration due to the sink's removal of defects. This reduction in vacancy concentration likely contributes to the lack of oxide formation in the region near the grain boundary as vacancies facilitate solute diffusion that leads to oxide formation.

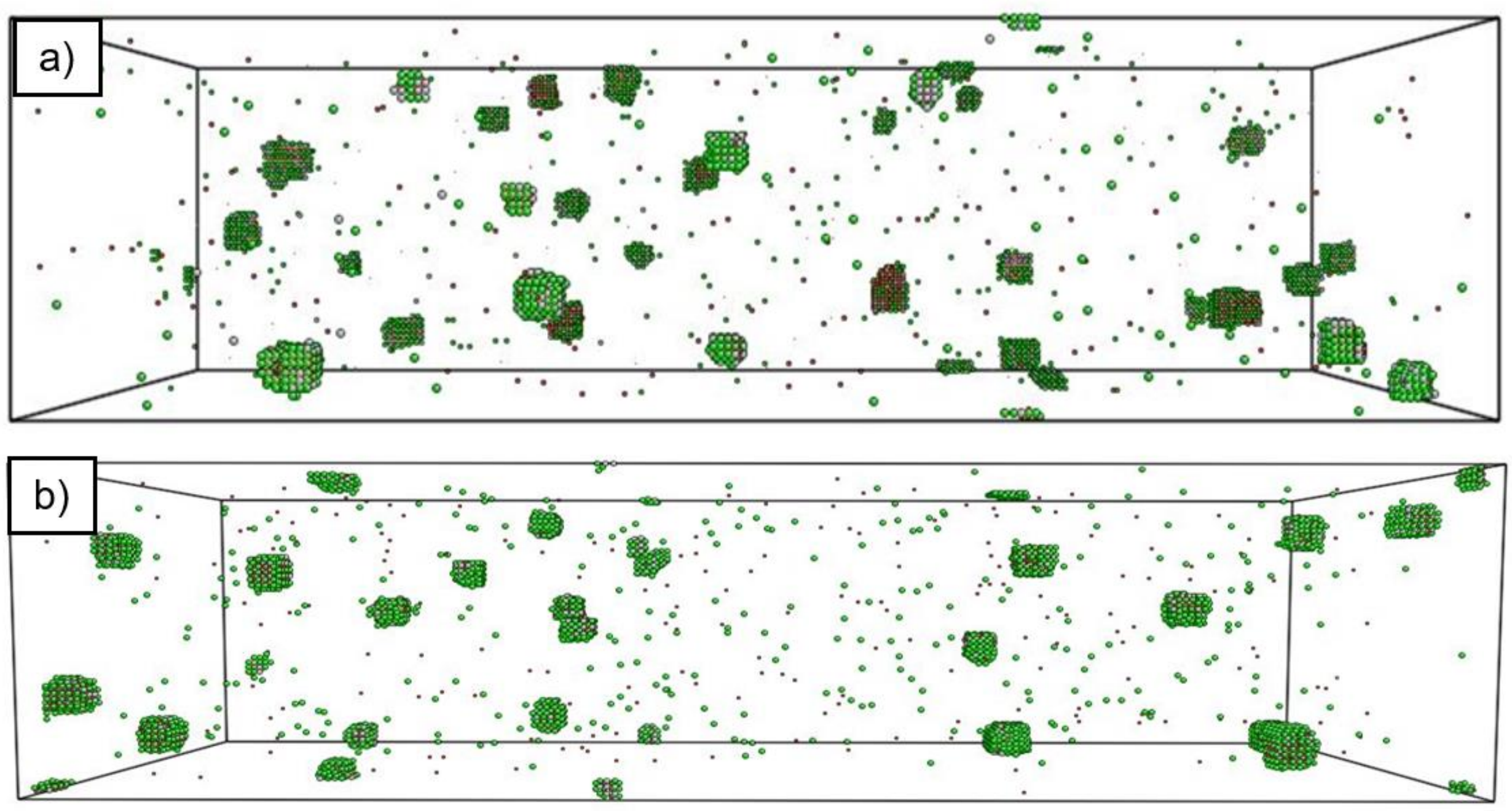

**Figure 3:** Oxide distributions for 12YWT a) in bulk b) with a grain boundary in the center of the simulation box. The size of the simulation box is 400x100x100 lattice sites.

3.1.2. Oxide Composition and Vacancy Concentration

The KMC model was further validated by comparing the composition of the oxides formed during the simulation against experimental findings. Tables 4 and 5 display the oxide compositions from the precipitation heat treatments in systems with and without the grain boundary as well as their expected compositions based on the literature. Both sets of simulations yielded similar oxide compositions, indicating that the defect sink had minimal influence on the composition of oxides outside its immediate vicinity. The oxide compositions have an approximate 1:1 ratio of metal to oxygen atoms (M:O), with more Ti atoms than Y atoms within the oxide. This ratio reasonably aligns with the experimental oxide composition, although the minor constituent Cr was absent from the oxide compositions due to its exclusion from the KMC model. The deviations from the stoichiometry of the oxide's chemical formula are attributed to the unbalanced starting concentrations of solutes.

After thermal aging, the vast majority of the Y, Ti, and O atoms became associated with oxide precipitates, and the concentration of solutes in the Fe matrix matched the solubility product in iron. The cubical shape of the oxide precipitates, as shown in Fig. 3, tends to align with the expectations set by Ribis [38].

**Table 4:** The atomic compositions of the Y-Ti-O oxides from the 12YWT -1123K heat treatment with and without a grain boundary in the simulation box and an experimental comparison.

| Material System | Y at% | Ti at% | O at% |
|---|---|---|---|
| 12-YWT KMC without GB | 12.9% | 37.6% | 47.9% |
| 12-YWT KMC with GB | 12.7% | 36.1% | 50.9% |
| 12-YWT Alinger with GB[7] | 14.8% | 33.6% | 39.4% |

**Table 5:** The atomic compositions of the Y-Ti-O oxides from the 14YWT -1123K heat treatment with and without a grain boundary.

| Material System | Y at% | Ti at% | O at% |
|---|---|---|---|
| 14YWT KMC without GB | 12.1% | 36.9% | 50.8% |
| 14YWT KMC with GB | 12.0% | 36.4% | 51.6% |
| 14YWT Miller with GB[6] | $7.5 \pm 4.3$ | $42.2 \pm 5.6$ | $43.3 \pm 5.3$ |

### 3.2. Precipitation Results under Irradiation

#### 3.2.1. 12YWT Irradiation Simulations

This study utilized the final configuration from the 1123 K precipitation heat treatment for neutron irradiation simulations. The oxide characteristics for the 12YWT-1123K simulation were a number density of $2.0 \times 10^{24}$ m$^{-3}$ with an average radius of 0.8 nm. The simulations were designed to replicate reactor environments within the same ranges as those collected by Wharry for their review [11]. These irradiation simulations explored the influence of temperature and dose rate on oxide characteristics by varying the irradiation conditions for each set of samples. Three dose rates were chosen for this investigation: $10^{-3}$ dpa/s to determine the resistance of the oxides to dissolution under very heavy irradiation, $10^{-5}$ dpa/s to investigate the results at a dose rate closer to experimental rates, and $10^{-4}$ dpa/s to observe any patterns related to dose rate. Three irradiation temperatures, 673 K, 773 K, and 873 K, were evaluated to study the temperature dependence of oxide survivability.

Figure 4 depicts the average oxide radius as a function of accumulated damage in dpa for all tested temperature and dose rate combinations. Across all irradiation regimes, a slight reduction in oxide size was seen at the 1-4 dpa range before appearing to stabilize in the 4-8 dpa range. Additionally, the oxides exhibited reduced atom loss at lower temperatures. At the lowest dose rate ($10^{-5}\ dpa/s$), there was a higher reduction in oxide size compared to the oxides irradiated at the same temperature but at higher dose rates. However, variations in precipitate size between dose rates declined with decreasing temperature, with the 673 K samples displaying minimal variability in results across different dose rates.

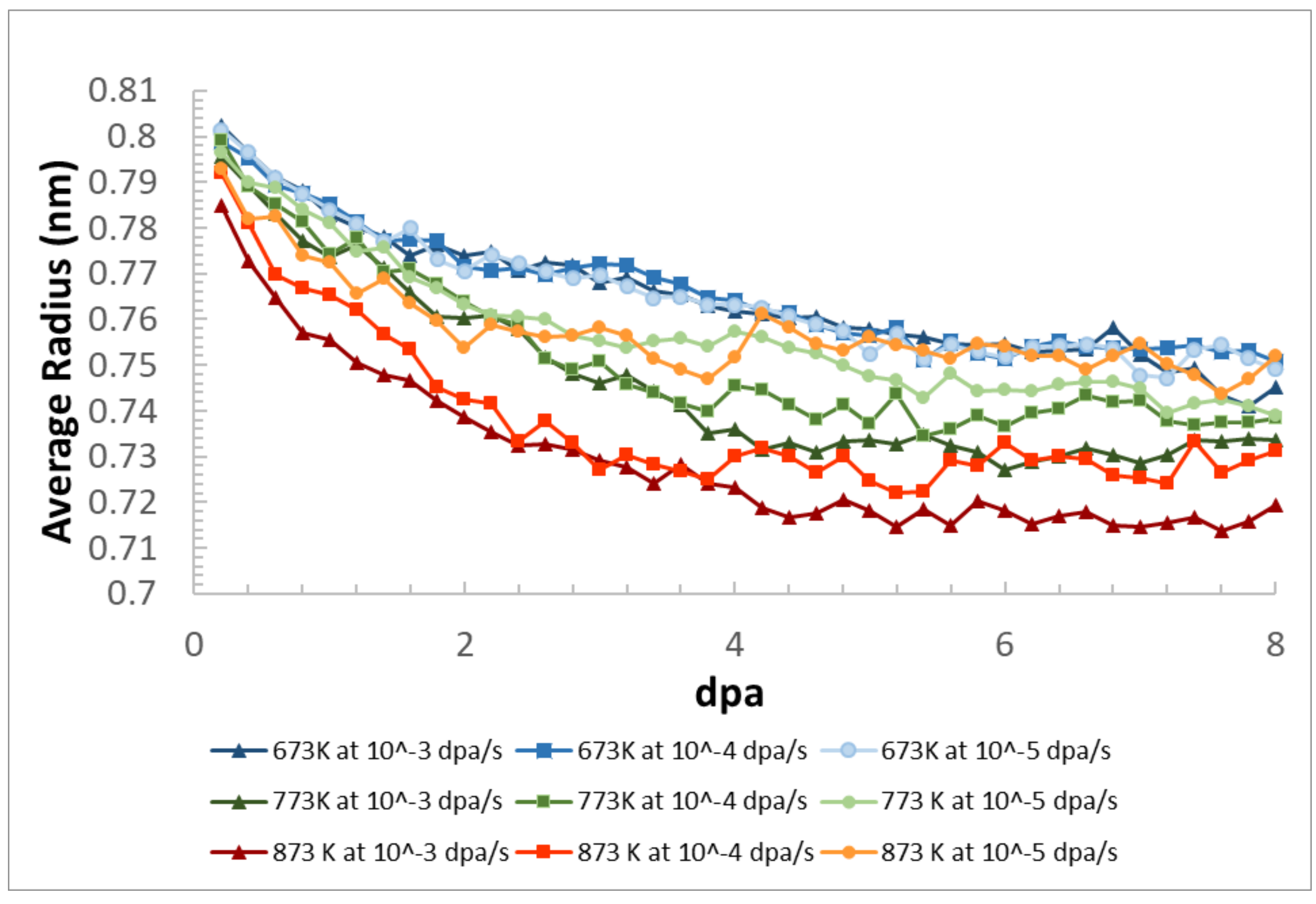

**Figure 4:** Evolution of the average oxide size in 12YWT over total accumulated dose at a variety of irradiation regimes.

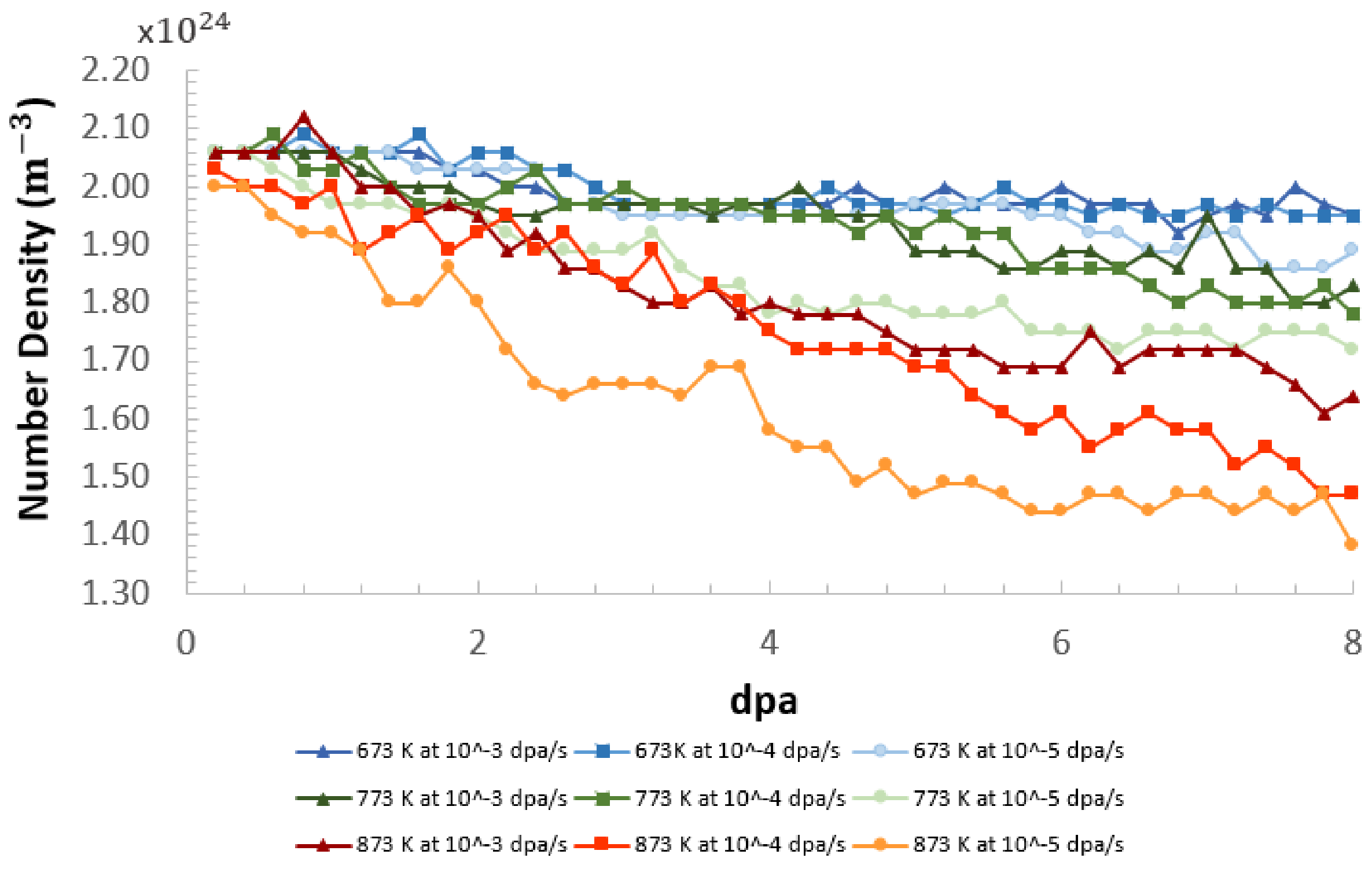

**Figure 5:** Evolution of the number density of oxides in 12YWT over total accumulated dose at a variety of irradiation regimes.

Figure 5 displays the change in the number density of oxides in the 12YWT as a function of accumulated dose in dpa for all tested temperature and dose rate combinations. Similar to the observations regarding average size, there was an abrupt initial drop in the number density before reaching a seemingly stable phase in the 4-8 dpa range. This reduction in oxide number density is consistent with findings in other neutron-irradiated NFA samples [13]. Higher temperatures led to increased oxide dissolution, with the most significant dissolution occurring at 873 K. In contrast, there was only a 10% reduction in the number density at 673 K and 773 K. Furthermore, more oxide dissolution was observed at lower dose rates than at higher dose rates across all tested temperatures, with the $10^{-5}$ dpa/s case at 873 K experiencing the most substantial decline with nearly a 30% loss in oxide number density.

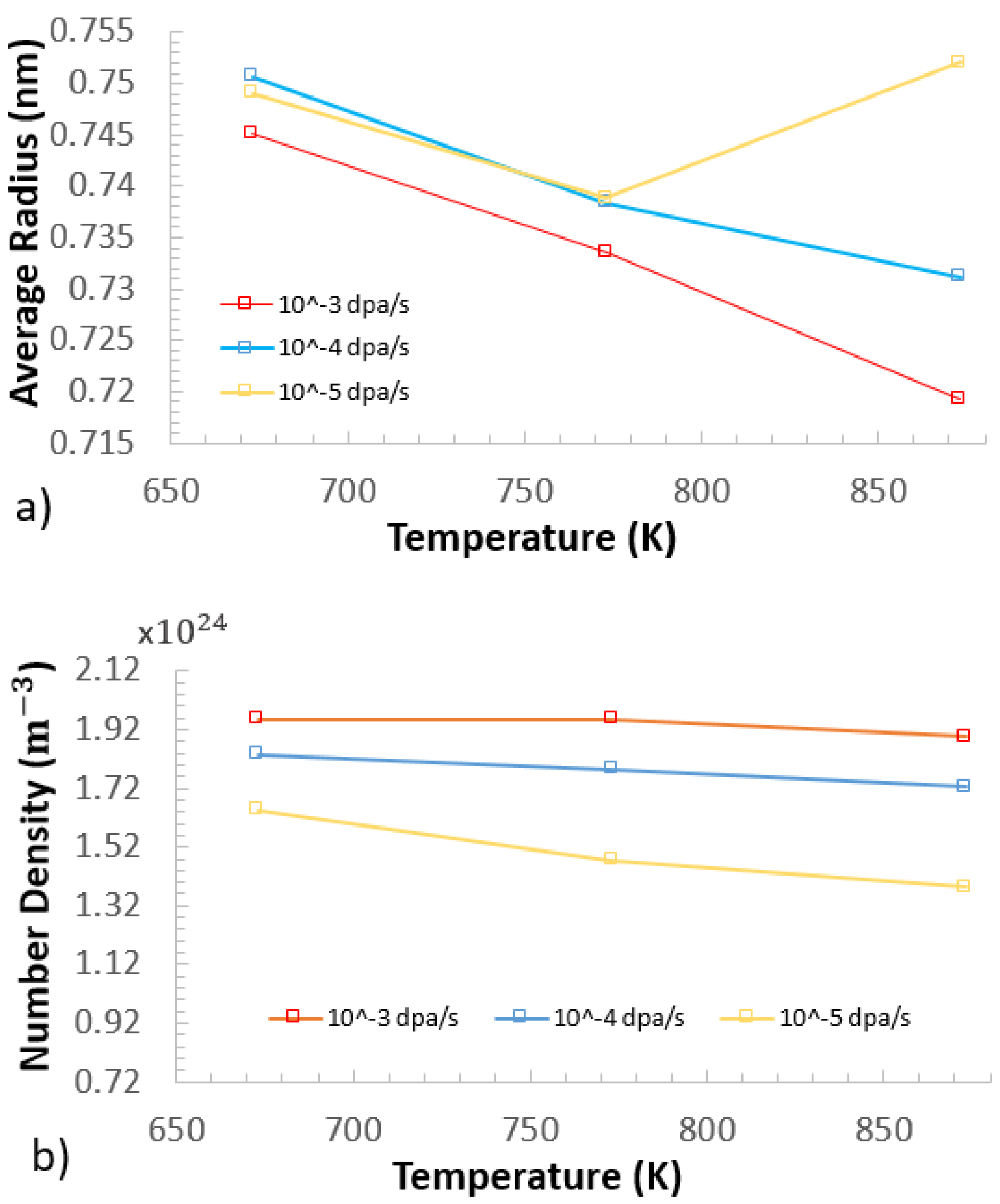

**Figure 6:** a) Average radius and b) number density of the nano-oxides in 12YWT-1123 K of a particular dose versus irradiation temperature.

Figure 6 illustrates the influence of temperature and dose rate on the average size and number density of the oxides at 8 dpa. As expected, the average radius decreases with higher irradiation temperatures. However, an interesting finding is that nano-oxides exhibit greater stability at higher rather than lower dose rates. This outcome is likely due to the increased point defect population at higher dose rates. These vacancies transport solute material back to the oxide precipitates more rapidly, mitigating their atomic loss. The reduction in average size at 873 K can also be linked to lower vacancy concentration as the higher temperature due to increased rates of recombination and annihilation at grain boundaries from faster diffusion, so the oxides are not adequately replenished with solutes delivered via vacancy diffusion at a sustainable rate than at 673 K and 773 K. Ribis's research on ion irradiation of ODS steels emphasized the crucial role of vacancy concentrations in the stability of nano-oxides. The temperature dependency of this factor can explain the discrepancies in nano-oxide irradiation stability observed at various temperatures [39].

The shape of the precipitates was visually inspected during thermal aging and neutron irradiation. In thermal aging simulations, a large, perfectly cubic Y-Ti-O oxide is aged within the simulation until it reaches an equilibrium shape.

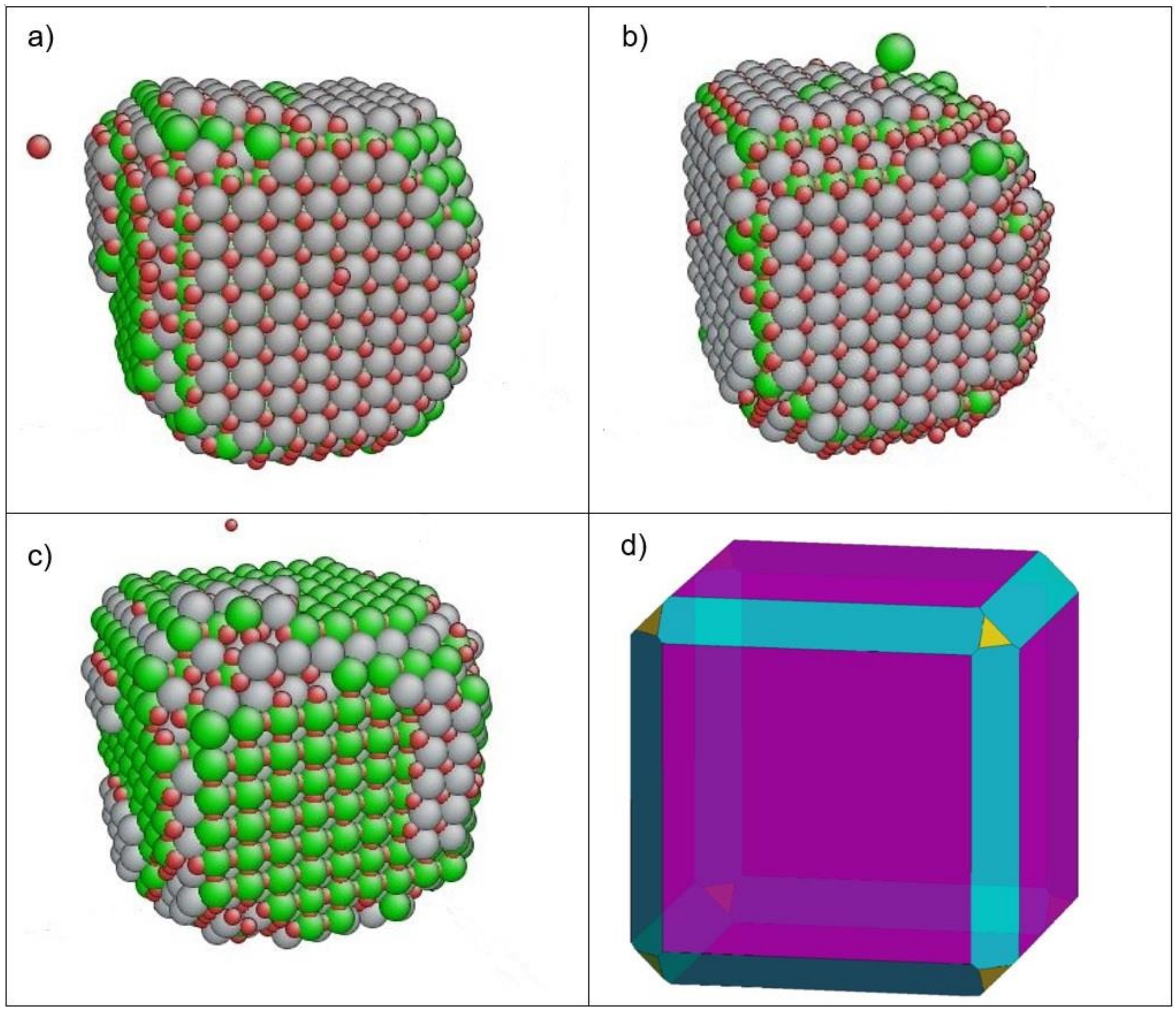

**Figure 7:** Equilibrium shape of the Y-Ti-O oxide at a) 673 K b) 773 K c) 873 K. Precipitates are compared with the expected equilibrium shape d) of the nano-oxides made using Wulffmaker [40]. Purple = <100>. Blue = <110>, Yellow=<111>.

Figure 7 displays the precipitates that were thermally aged at all three irradiation temperatures (673 K, 773 K, 873 K). The theoretical equilibrium shape for the oxide is shown in Figure 7d and is determined based on the interface energies presented in Table 6 using the Wulffmaker [40] program. The oxide precipitates were primarily characterized by the <100> interface, with smaller areas of the <110> and <111> interfaces. Both the simulation and theoretical model align closely, successfully reproducing the experimental shape observed by Ribis [38]. The oxides in the irradiation simulations continued to maintain this shape as shown in Figure 8. Figure 8 presents an image of an oxide in the 12YWT-1123 K system after irradiation at 773 K with a dose rate of $10^{-3}$ dpa/s, resulting in a total dose of 8 dpa. This image is representative of the changes in oxide shape observed in all the simulations. As irradiation increased, these oxide faces became more diffuse and irregular, while still maintaining a cubic/rectangular shape, consistent with expectations [41].

**Table 6:** Interface energies from the broken bond method (BBM).

| Interface Energy | Y-Ti-O (BBM) |
|---|---|
| $E_{100}^{\gamma}$ | 0.41 |
| $E_{110}^{\gamma}$ | 0.52 |
| $E_{111}^{\gamma}$ | 0.63 |

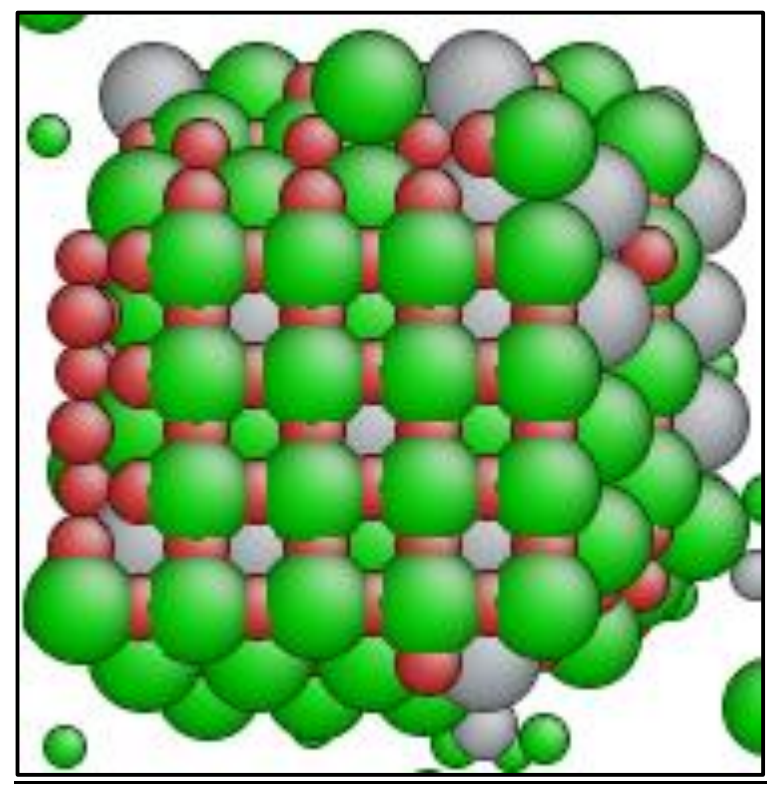

**Figure 8:** 12YWT-1123 K after irradiation at 773 K at a dose rate of $10^{-3}$dpa/s to a total dose of 8 dpa. Green spheres: Ti atoms, Grey spheres: Y atoms, Red spheres: O atoms.

3.2.2. 14YWT Irradiation Replication Simulations

After completing the irradiation simulations for the 12YWT-1123 K samples at the three test temperatures and three dose rates, the investigation into the general response of the oxides in 14YWT-1123 K to irradiation was carried out. Figures 9 and 10 show the change in average radius and number density as a function of accumulated damage in 6 simulations of 14YWT, using the conditions described in Aydogan [42]. There was a decrease in the average radius that appeared to be linear with the accumulated dpa damage. The reduction in nano-oxide size under irradiation at these temperatures was consistent with the findings of many other experimental studies [13]. Unlike the 12YWT irradiation simulations, a plateau in oxide size reduction did not appear to have been reached in this case. The number density of the oxides remained constant during irradiation across all samples.

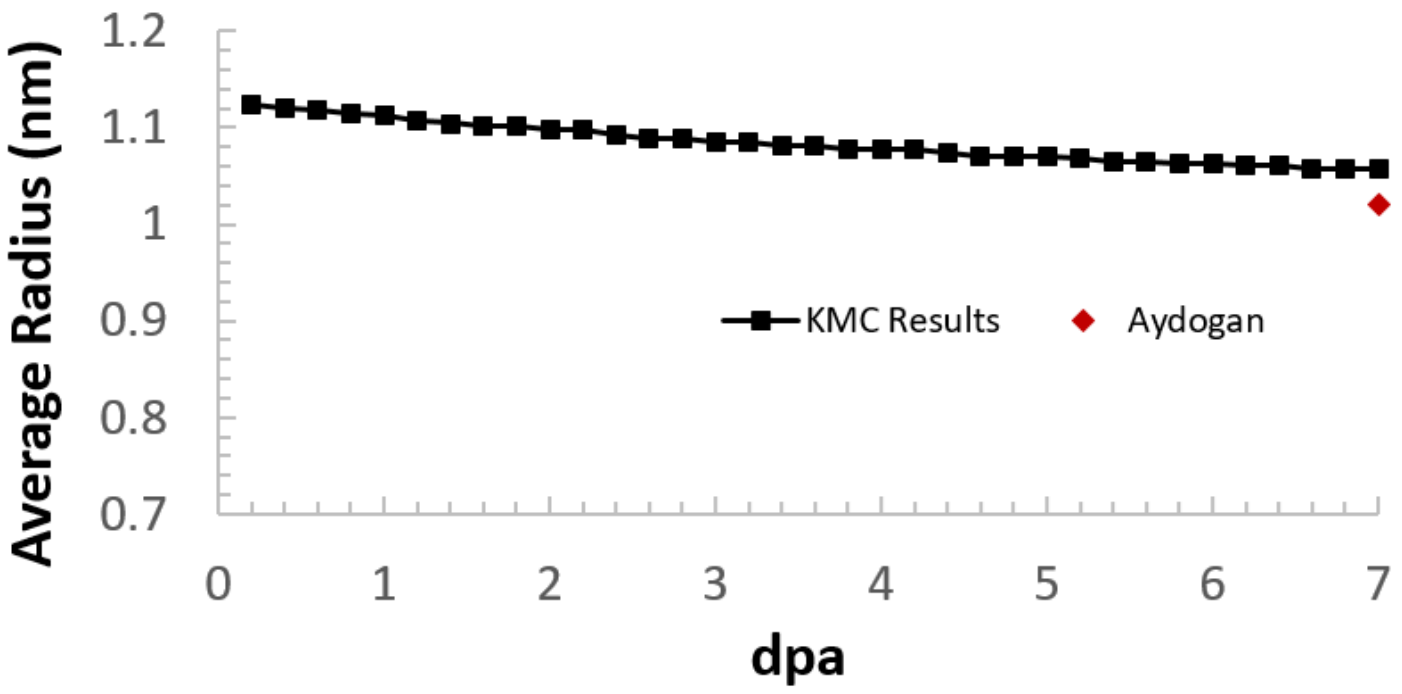

**Figure 9:** Average radius (nm) over dpa for the 14YWT-1123K irradiation at 633K. Red dot represents Aydogan's observations [42].

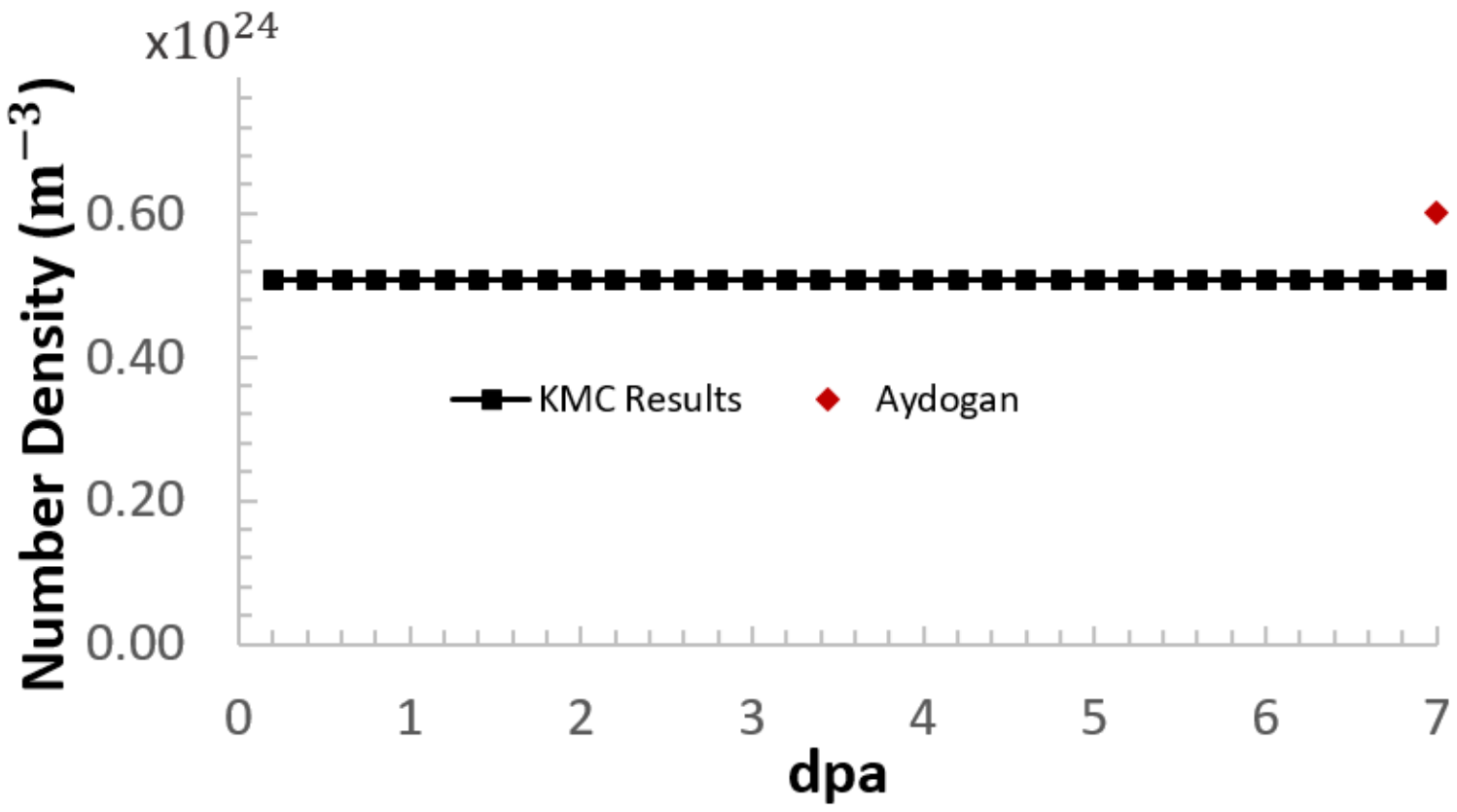

**Figure 10:** Number density over dpa for the 14YWT-1123 K irradiation at 633K. Red dot represents Aydogan's observations [42].

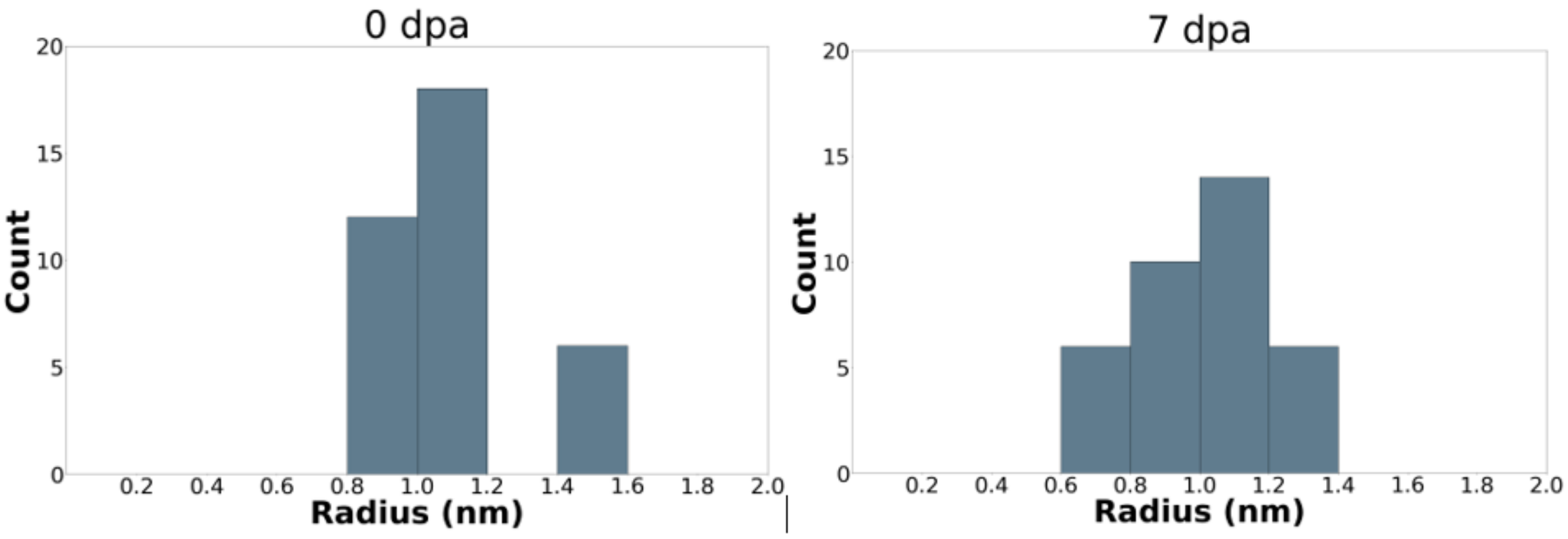

**Figure 11:** Oxide size distributions of 14YWT before and after irradiation to 7 dpa.

Figure 11 displays the size distribution of the oxides across all 6 simulations before and after irradiation to 7 dpa. None of the oxides fully dissolved though all of them experienced a decrease in size, with no irradiation-induced growth observed, which aligns with the average size and number density trends seen in Figures 9 and 10.

3.2.3. Composition Change of Precipitates after Irradiation

Shifts in the composition of the oxides in the NFA due to irradiation could affect their performance and stability. The composition of the oxide precipitates after the irradiation treatment was compared to the values at the beginning of the simulation for the 12YWT and 14YWT compositions. Tables 7 and 8 show the composition of the oxides after irradiation and their respective compositions prior to irradiation. In both cases, the proportion of O and Fe in the oxide increases, while the proportion of Y and Ti in the oxides decreases. The decrease in Y concentration with increasing temperature matches well with other irradiation damage experiments [39]. There is no consistent variation in oxide composition with dose rate, with the composition being more strongly dependent on temperature. The proportion of Ti in the oxide declined the most over the course of the irradiation, particularly at higher irradiation temperatures. Nearly a third of Ti atoms in the oxide were expelled at 873 K. This high reduction in Ti concentration due to neutron irradiation is observed in other experimental results [43]. One of the reasons is the higher diffusion coefficient of Ti compared to Y or Fe.

**Table 7:** Composition of oxides after irradiation to 7 dpa at $6.5 \times 10^{-7}$dpa/s compared to literature values.

| | **O at%** | **Y at%** | **Ti at%** | **Fe at%** |
|---|---|---|---|---|
| 14YWT from KMC at 0 dpa | 51% | 12.0% | 36% | - |
| 14YWT KMC After Irradiation | 54.1% | 9.8% | 28.85% | 7.14% |
| 14YWT Miller[6] | $43.3 \pm 5.3$ | $7.5 \pm 4.3$ | $42.2 \pm 5.6$ | - |

**Table 8:** Composition of oxides in 12YWT after irradiation to 8 dpa per irradiation regime.

| | **Dose Rate (dpa/s)** | **O at%** | **Y at%** | **Ti at%** | **Fe at%** |
|---|---|---|---|---|---|
| **12YWT Before** | N/A | 50.94 | 12.76 | 36.17 | 0.17 |
| **673 K** | $10^{-3}$ | 56.10 | 10.23 | 28.18 | 5.48 |
| **673 K** | $10^{-4}$ | 55.57 | 10.45 | 28.48 | 5.49 |
| **673 K** | $10^{-5}$ | 55.58 | 10.14 | 27.77 | 6.27 |
| **773 K** | $10^{-3}$ | 58.61 | 9.71 | 26.08 | 5.59 |
| **773 K** | $10^{-4}$ | 58.6 | 9.51 | 26.34 | 5.51 |
| **773 K** | $10^{-5}$ | 58.789 | 9.53 | 25.63 | 6.04 |
| **873 K** | $10^{-3}$ | 61.58 | 9.14 | 22.45 | 6.81 |
| **873 K** | $10^{-4}$ | 62.44 | 8.88 | 22.27 | 6.39 |
| **873 K** | $10^{-5}$ | 62.35 | 9.00 | 22.38 | 6.25 |

Figure 12 shows the composition evolution during the irradiation simulations. As the total accumulated dose increased, more Y and Ti solutes were ejected from the oxides into the bulk. Fe was gradually introduced as a replacement component in the oxide, increasing from near non-existent to up to 7 at%. In all cases, composition changes appeared to be heading towards a plateau near the end of the simulation, suggesting that the oxide composition would reach a steady state value where solutes are replaced as quickly as they are ejected from the precipitates. This visualization demonstrates that the composition was more strongly influenced by the irradiation temperature than the dose rate.

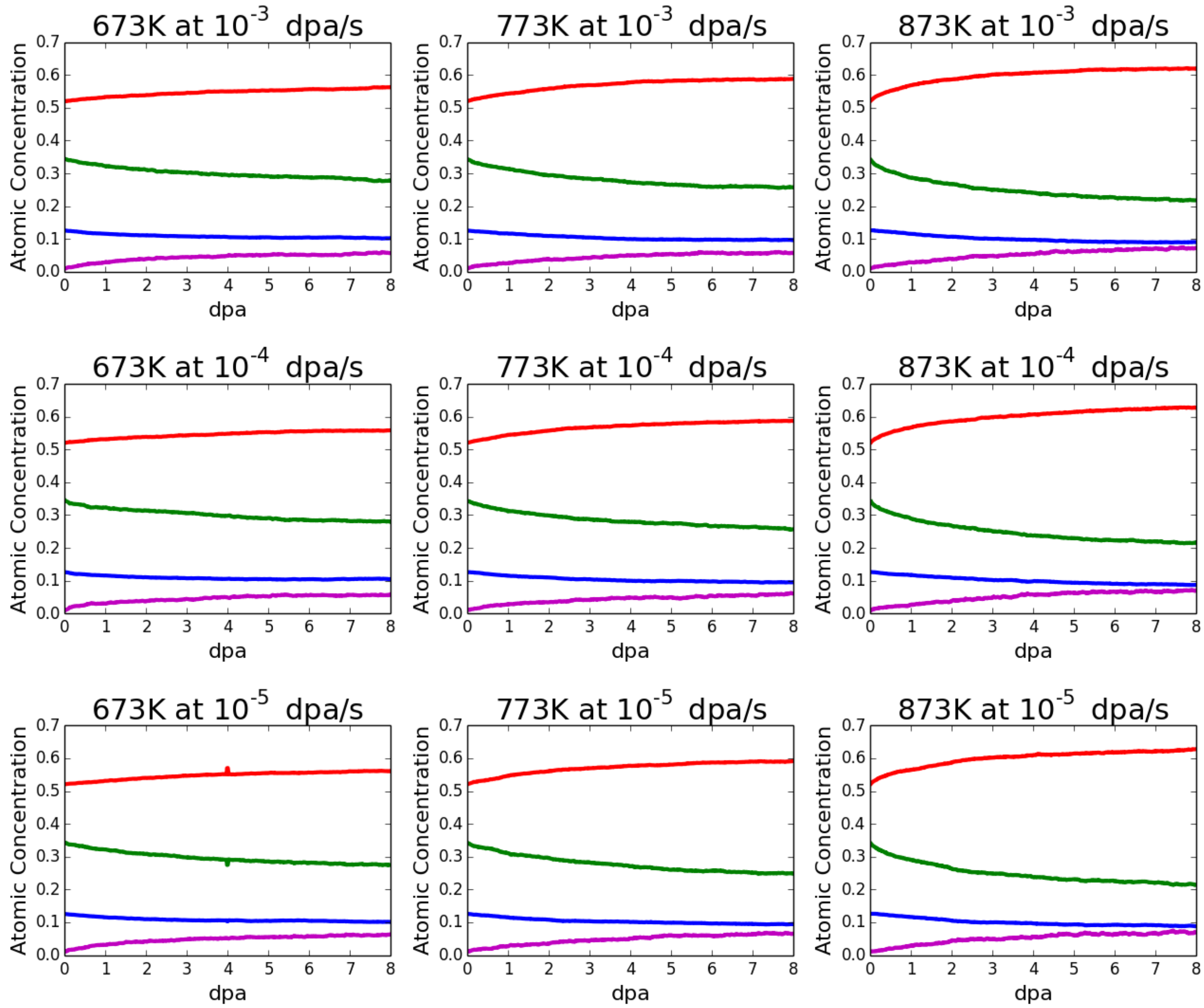

**Figure 12:** Evolution of oxide composition in 12YWT over accumulated dose at various temperatures and dose rates. Red=Oxygen, Green=Titanium, Blue=Yttrium and Purple = Iron.

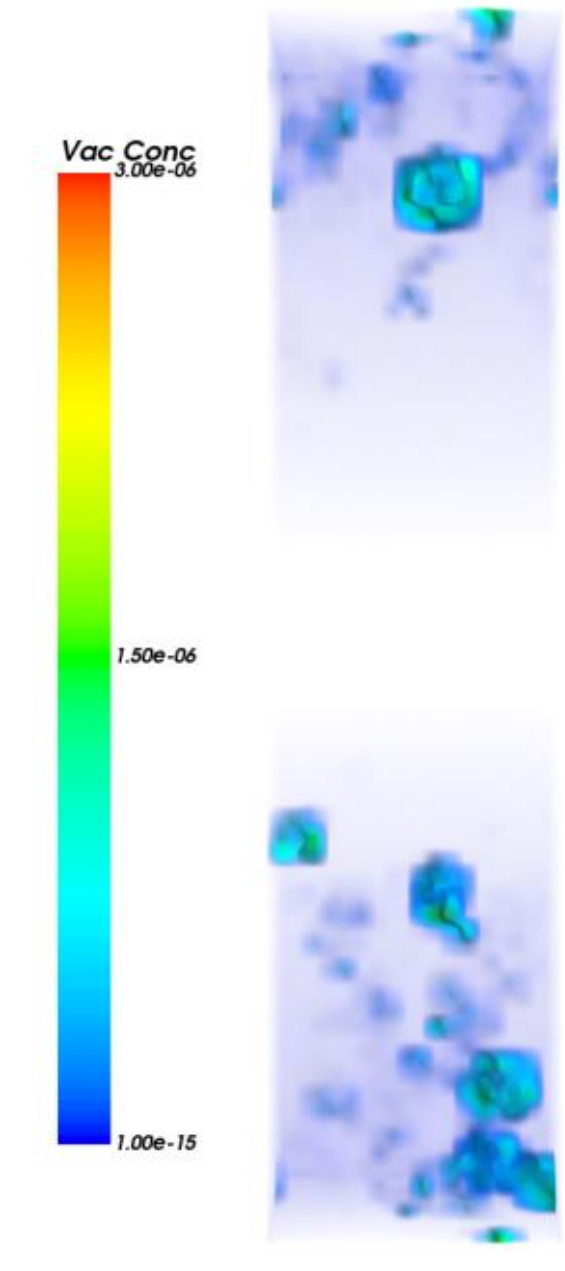

**Figure 13:** Average vacancy concentration heat map of the 14YWT under neutron irradiation at 773K and $10^{-3}$ dpa/s.

Figure 13 displays the vacancy heat map of the 14YWT under irradiation. Similar to the precipitation simulations, free vacancies spend more time at the oxide interfaces than in the bulk. This behavior is due to the attractive interactions between the oxide constituent atoms and the vacancy, in comparison to Fe. An earlier investigation found a binding energy between the vacancy and the oxide of around 1.1 eV [22]. Additionally, even under high irradiation, there is still a lower vacancy concentration in the grain boundary region than in the outer regions, due to the grain boundary's role as a defect sink.

3.2.4. High DPA Regime

Since the NFAs are expected to receive a lifetime dose in the 100 dpa range, an investigation into the stability of the nano-oxides under longer-term irradiation was conducted. After observing an apparent plateau in the decline of the size and number of oxides in the system across all irradiation regimes, an additional limited investigation was carried out to see if that plateau represented the emergence of a steady state or only a temporary metastable stage. Since it is currently impractical to extend the simulated irradiation regimes past 8 dpa for all conditions, the single regime selected for long-term irradiation was chosen primarily for its real-time speed of completion in comparison to the other regimes. In this case, the irradiation regime at 873 K and $10^{-3}$ dpa/s was used to expose three 12YWT samples to a total dose of 66 dpa. The evolution of the oxide characteristics as a function of dose was recorded for analysis.

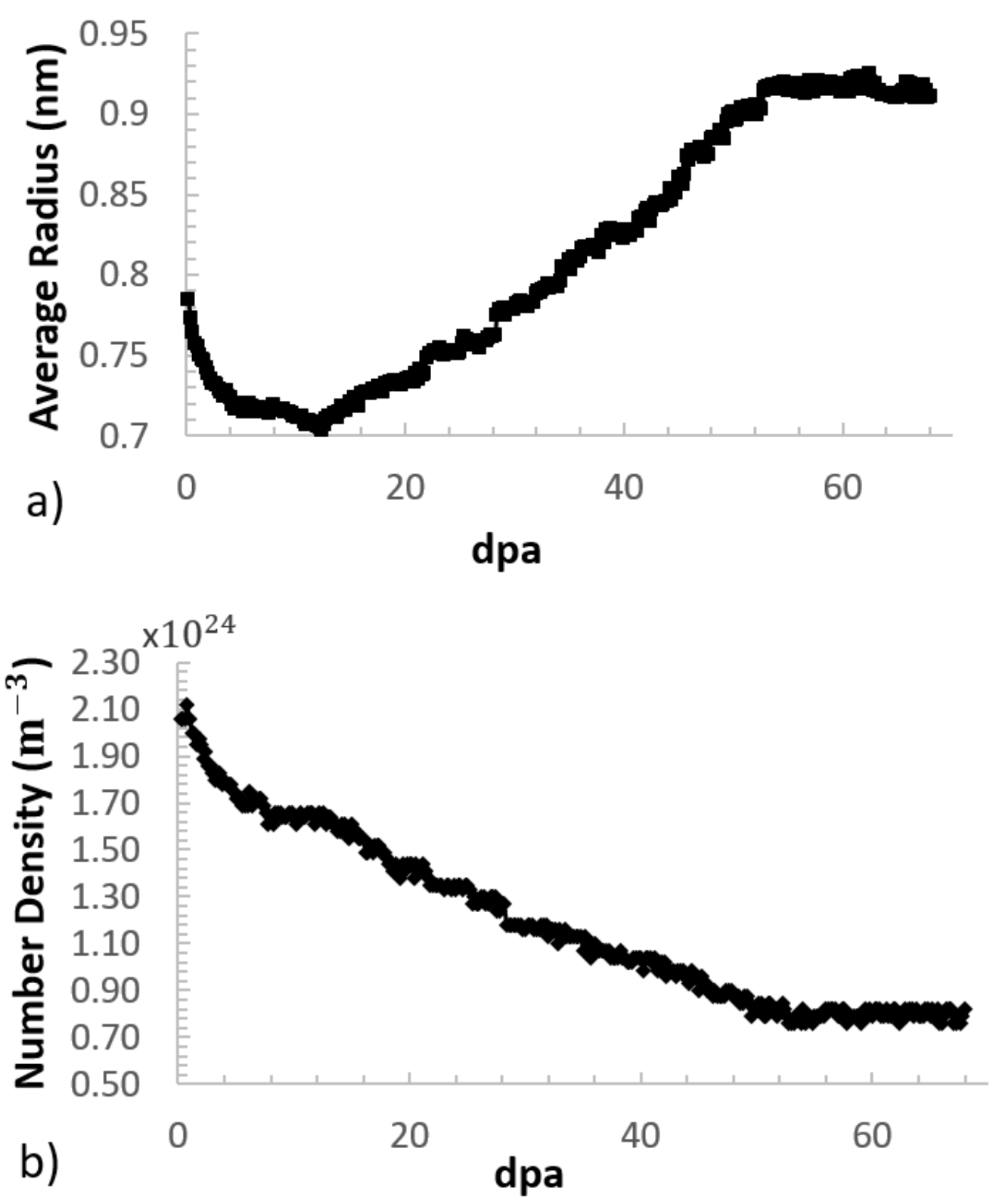

**Figure 14:** The evolution of 12YWT oxide characteristics in long-term irradiation simulation at a single irradiation regime (873K at $10^{-3}$ dpa/s).

Figure 14 shows the evolution in oxide size and number density as they were irradiated to a total dose of 66 dpa at 873K at $10^{-3}$ dpa/s. After 8 dpa, the apparent plateau ceased and the oxide size began to decline again before reversing and increasing in size after 12 dpa. The change in the number density of oxides continued to follow its initial downward trend throughout the entire simulation and no longer mirrored the change in oxide size. The evolution reaches an apparent steady state after 54 dpa, characterized by no loss of oxides and no change in the average radius. While solutes are still being ejected from the surviving oxides at this high dpa, there are now enough solutes in the Fe matrix from oxide dissolution to replace those ejected solutes at roughly the same rate. These results align with experimental findings of neutron irradiation damage at high dpa regimes, which exhibit an increased average radius and a decrease in number density [12]. A steady-state region after long-term irradiation has been predicted by Chen [44].

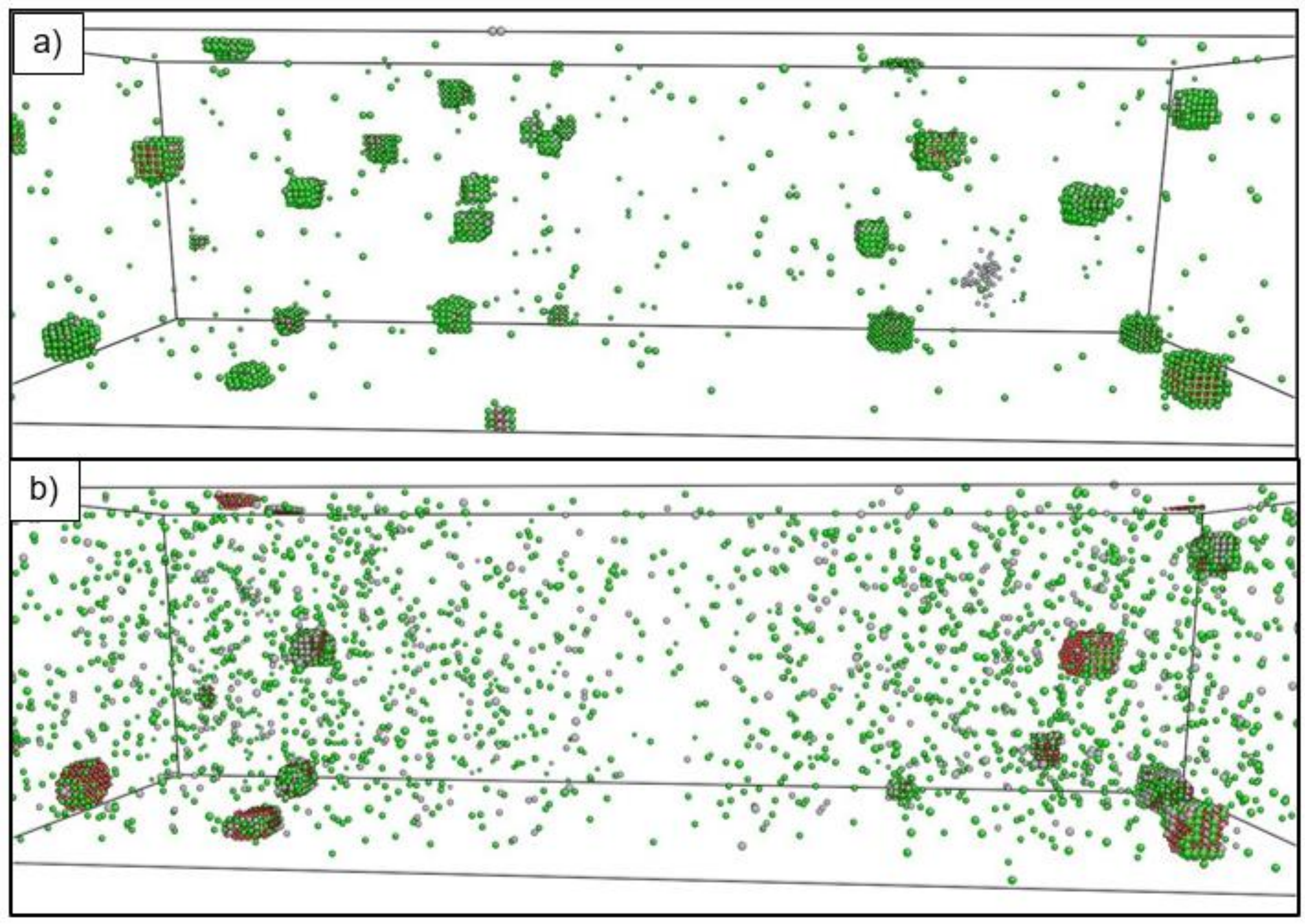

**Figure 15:** Oxide distributions of the 12YWT-1123K a) before neutron irradiation, b) after neutron irradiation with a total dose of 66 dpa at 873K at a dose rate of $10^{-3}$ dpa/s.

Figure 15 shows pictures of the 12YWT simulation box before and after neutron irradiation to 66 dpa. In the irradiated sample, there was a higher density of free solutes Ti and Y in the Fe matrix, separated from their original oxides due to irradiation. Many of the original oxides dissolved under irradiation, with oxides around the grain boundary suffering the most losses. The grain boundary itself was largely devoid of free solutes, and no new oxide precipitates were found at the grain boundary.

The tendency of a solute species to enrich or deplete at defect sinks is attributed to the inverse Kirkendall effect, where preferential transport of the solute via vacancy or dumbbell migration compared to matrix atomic species leads to segregation around defect sinks. Wiedersich [45] developed a formulation that predicts whether a solute enriches or depletes at defect sinks using the diffusivities of the solute and matrix atoms.

$$\left(\frac{d_{AV}}{d_{BV}} - \frac{d_{AI}}{d_{BI}}\right) \rightarrow \begin{cases} < 0 & Enrichs \\ > 0 & Depletes \end{cases} \tag{6}$$

The diffusivities $d_{XV}$and $d_{XI}$ of atomic species X via vacancy and dumbbell migration respectively. They are calculated using

$$d_{AV} = d_{AV}^{0} \exp(\frac{-E_{mig}}{k_b T}) \tag{7}$$

Where $d_{AV}^{0}$ is the pre-exponential factor, and $E_{mig}$ is the migration energy of solute A by the vacancy mechanism. The values for these parameters are taken from Table 2. The remaining diffusivities ($d_{AI}, d_{BV}, d_{BI}$) are also calculated using Eq. 7. To check the direction of segregation using Eq. 6, atomic species A is the solute (Y or Ti) and atomic species B is the matrix atom (Fe).

**Table 9:** The Ratio of Diffusivity of Y and Ti with Fe.

| $\left(\frac{d_{AV}}{d_{BV}} - \frac{d_{AI}}{d_{BI}}\right)$ | A=Y<br>B=Fe | A=Ti<br>B=Fe |
|---|---|---|
| 673K | -10.32 | 19.68 |
| 773K | -7.84 | 31.60 |
| 873K | -6.23 | 43.33 |

Table 9 highlights the diffusivity ratios for the Y and Ti solutes in bcc Fe at all irradiation temperatures. There was a slight preference for enrichment of Y atoms at defect sinks, while the Ti atom was more strongly inclined toward depletion. The presence of oxides and the supersaturation of Y and Ti solutes in the Fe matrix were likely the causes of the differences in predicted behavior for Y segregation, as Wiedersich's formulation assumes that these conditions are not present.

3.2.5 Point Defect Suppression

The nano-oxides are suspected of suppressing the point defect population and thereby slowing the diffusion processes that drive segregation. To investigate this suspected phenomenon, the KMC simulations were run at three temperatures (673 K, 773 K, and 873 K) at a single dose rate of $10^{-3}$ dpa/s for two starting configurations. The first configuration was pure Fe with no other solutes, and the second was the 14YWT configuration. The average concentration of point defects in the bulk matrix (not bounded to the oxides) was recorded and compared between the two cases.

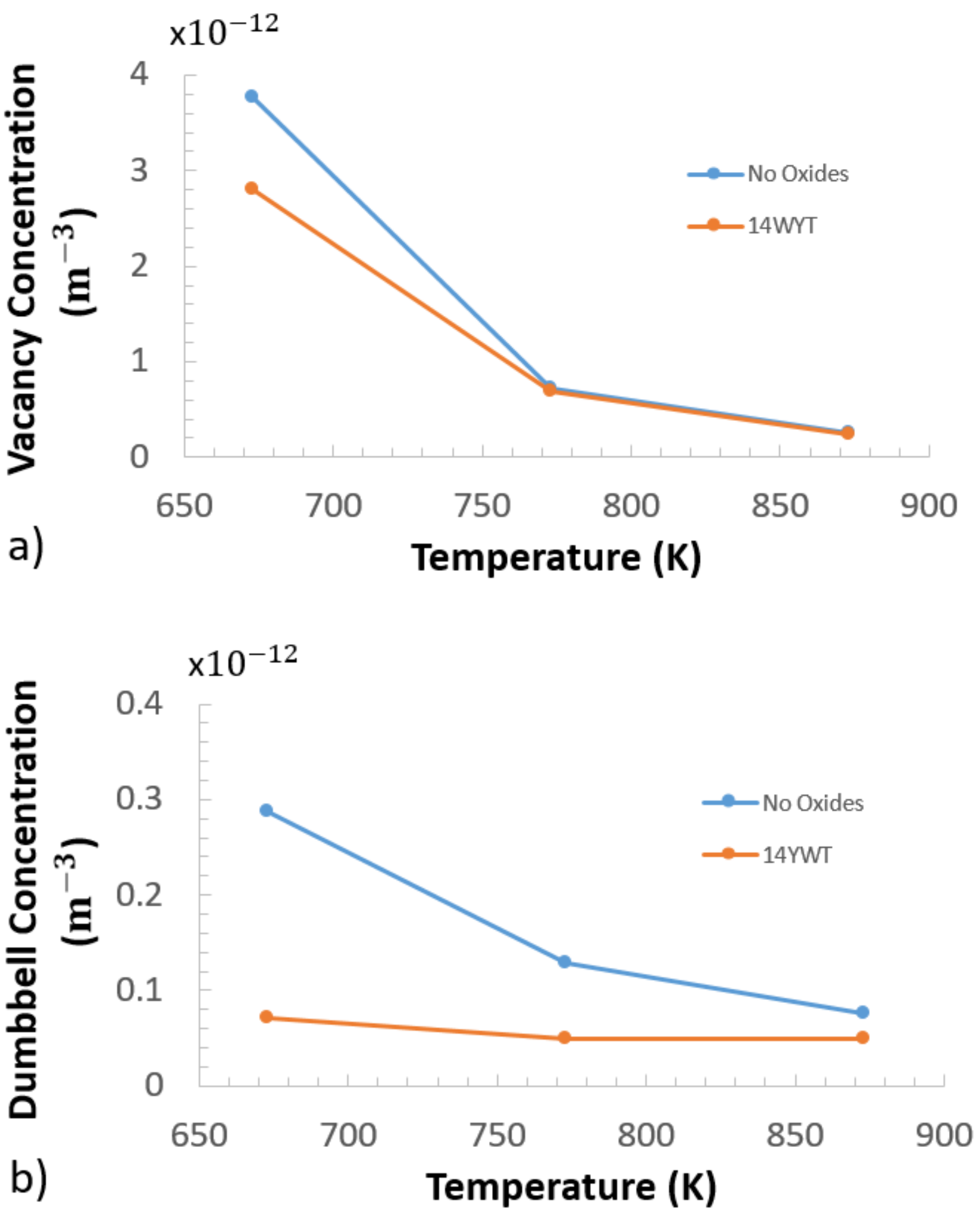

**Figure 16:** The concentration of point defects in pure Fe and 14YWT vs temperature. a) Vacancy concentration. b) Dumbbell concentration.

Figure 16a and 16b show the average concentration of vacancies and dumbbells, respectively, in the simulation box during irradiation at $10^{-3}$ dpa/s at different temperatures for both systems. The plots show that the nano-oxides in 14YWT do have a suppressing effect on the average population of point defects. This suppression effect weakens with increasing temperature. At 873 K, there is almost no difference in the concentration of vacancies. This outcome is likely due to the high temperature speeding up recombination and diffusion to the grain boundary, preventing the vacancy concentration from rising much above the thermal equilibrium vacancy concentration.

# 4. Discussion

## 4.1. Precipitation

The precipitation simulations showed a pattern of increasing oxide size with rising temperature due to the increased solubility product of the oxides at higher temperatures. This larger solubility product critical radius, making it less likely for stable dispersoid nuclei to form through thermal processes. Consequently, free solutes tend to aggregate at the fewer stable nuclei, leading to a lower number density of larger-sized oxides. The size, number density, and composition of the oxides in both 12YWT and 14YWT simulations compared favorably to experimental results. Any discrepancies observed during the precipitation heat treatments may be influenced by the absence of solutes such as Cr in the KMC model. Regarding the shape of the precipitates, the oxide precipitates in the precipitation heat treatments took on a cubic shape, characterized by broad <100> interfaces and smaller areas of <110> and <111> interfaces at higher temperatures. This cubic shape of the oxide precipitates aligns with the equilibrium shape predictions derived from the interface energies and corresponds to observations from experiments.

The study investigated the impact of the grain boundary on precipitation at the same temperatures as a prior paper that examined the precipitation of Y-Ti-O oxides in bulk Fe. The results indicated minimal differences in the average oxide radius and number density between the bulk and grain boundary simulations. However, there was a distinction in oxide distribution between the two scenarios, which was the absence of oxide precipitates forming in the immediate vicinity of the planar sink. There is literature that reported the formation of oxides at grain boundaries. These grain boundary nano-oxides were typically larger than the oxides in the bulk region. One theory suggests that the presence of interstitial elements such as C and N may facilitate the formation of nano-oxides at grain boundaries. Thus, the absence of these elements in our simulations could explain the absence of grain boundary nano-oxides in our results.

Another factor to consider is the opportunity for vacancy diffusion at the grain boundary. Due to the sink's annihilation of point defects upon contact, this region had a significantly lower average concentration of vacancies compared to areas away from the boundary. Thus, there are fewer opportunities for atomic diffusion around the grain boundary, which inhibited the nucleation of Y-Ti-O precipitates. Given NFA's known rapid oxide nucleation, it is likely that the matrix would likely be depleted of free solute material before any nucleation near the grain boundary could take place. Furthermore, thermal segregation of intergranular Ti, Y, and O atoms to the grain boundary during heat treatments was not observed in Alinger's work [5], which means there is less available material grain boundary nano-oxides.

### 4.2. Irradiation

#### 4.2.1. Overall Survivability

In the neutron irradiation simulations, the behavior of the oxides aligned with our expectations in response to irradiation. The simulations found that the oxides behaved as expected, showing resistance to dissolution even at the relatively high dpa rate of $10^{-3}$ dpa/s with the high dose rate appearing less effective at dissolving oxides. The presence of vacancies worked to heal the oxide structures when the displacement cascade ejected solutes from the oxides. Nevertheless, there was still a decrease in the overall size of the oxides over the irradiation regime. Examining the size distributions before and after the irradiation showed that the smallest oxides quickly dissolved, while the larger oxides were more stable. For the 12YWT cases at accumulated damage > 4 dpa, there were no oxides larger than 1 nm in radius. The stability of the oxides could also be influenced by the inherent KMC setup. Chen found that oxides coherent with the Fe matrix were more stable than incoherent oxides, implying that the oxides represented in the rigid lattice KMC setup might be more stable than in reality [44].

The inherent stability could also be explained by Nelson-Hudson-Mazey’s theory of stability, which depends on two competing processes induced by irradiation [46]. One is where irradiation expels debris from oxides and the other is the radiation-enhanced diffusion that drives the regrowth of the oxides. Similar to how thermal vacancies drive nucleation and growth during heat treatment, the vacancies produced by irradiation drives a ‘healing’ effect on the damaged oxides by delivering solutes from the matrix to the oxides. Additionally, the preferential transport of certain solutes through SIAs or dumbbell migration toward or away from the grain boundary or oxides plays a role in the reordering of oxides. As a result, the low dumbbell concentration at high temperatures also contributes to the reduced oxide survivability under irradiation. Therefore, more atomic loss in oxides is observed when the defect population is kept to a relative minimum at 873 K. Smaller oxides proportionally lost more material in the irradiation cascade, making them more susceptible to dissolution.

The long-term (> 66 dpa) irradiation simulations at 873 K and $10^{-3}$ dpa/s saw a reversal in the decline of oxide size, with the surviving oxides growing larger than the original average. Atomic material for this growth comes from oxides dissolved during irradiation, resulting in a decrease in the oxide number density. This radiation-induced coarsening of the oxides is observed in similar NFAs [12].

After irradiation, a compositional analysis of the oxides for both 12YWT and 14YWT found an overall decrease in the proportion of Ti atoms in the oxide with a corresponding increase in the proportion of Fe atoms in the oxide. Ti atoms constituted most of the solute atoms in the matrix. Due to the stronger binding of Y atoms to oxygen compared to Ti, Y atoms were less likely to separate from the oxide and were more likely to rejoin the oxide after being ejected. Overall, across the sampled temperatures, the oxide compositions generally approached a stable plateau after irradiation to 8 dpa.

#### 4.2.2. Precipitate Shape

During the heat treatments, the oxides precipitated into cubical shapes, which aligned with our expectations and provided another measure of validation for the KMC model. The shape of a precipitate is determined by the differences in energies at the oxide/matrix interfaces at various orientations. If the energetics of the system are set up properly, then the oxides formed will match experimental observations, which was found to be the case across all simulations. The shape was orderly with sharp interfaces that became more diffuse upon irradiation, owing to the constant dislodging of atoms by irradiation. These interfaces would likely become orderly again when the irradiation ceased.

#### 4.2.3. Effect of the Dose Rate

When examining the relationship between dose rate and oxide survivability, the following pattern became apparent: higher dose rates resulted in a larger reduction in oxide size, while at the lowest dose rate, there was less of a decrease in oxide size. However, at the lower dose rates, $10^{-4}$and $10^{-5}$ dpa/s, the difference in oxide size was minimal. The lower dose rates led to a smaller supersaturation of point defects, which slowed the diffusion that heals the oxides. As a result, the oxides experienced more atomic loss after 8 dpa. In all of the irradiation simulations with the 12YWT, there was a plateau reached, where the average size, number density, and composition did not change as quickly as before. This region is where enough solutes have been ejected into the matrix to provide a sufficient source of solutes for vacancy diffusion to start to stymy oxide material loss.

The limited long-term irradiation investigation, conducted to further explore the extent of oxide stability, found that the period of stability after 8 dpa was temporary, and oxide loss continued. Eventually, a point was reached where the oxide loss stabilized after 54 dpa, and the average size and number density of the oxides remained roughly constant until the end of the simulation at 66 dpa. This steady-state region, predicted by other researchers [11, 44], was where the smallest oxides dissolved completely, and the more stable larger oxides remained, balancing the loss and gain of atomic material in the oxides.

4.2.4. Segregation

In the irradiation simulations, there was a lack of segregation of the solute Y, Ti, and O atoms to the grain boundaries. There even appeared to be an inverse Kirkendall effect where the solutes were depleted in the region around the grain boundary. The lack of segregation of Y-Ti-O elements has been observed in experimental observations of irradiated NFAs, with an apparent preference for the solutes to return to the oxides [47]. Other studies have shown both segregation of Ti [48] and no segregation [49] to the grain boundary reported in some cases; however, these experiments involved heavy ion irradiations. The simplistic description of the grain boundary in the KMC model could have influenced the resulting segregation profiles. Field showed that Cr segregation could be highly dependent on grain boundary structure [50]. The oxides also suppress the point defect populations, thus partly stifling the degree of segregation.

4.2.5. Nano-oxide solute ejection and diffusion

From Wiedersich's formulation, Eq 6 predicts, based on the diffusivities of the solutes, that Ti would deplete and Y would enrich at defect sinks. While the tendency of Y to deplete is in contrast to this predicted Y behavior at defect sinks, factors unaccounted for in the formulation, such as the influence of oxides, should be considered. During irradiation, atoms are ejected from the oxides into the Fe matrix. Once in the matrix, there are several possible destinations for the atoms: 1) return to the original oxide, 2) join a preexisting oxide, 3) form a new nucleate with other ejected solutes, and 4) accumulate at the grain boundary and possibly precipitate a new phase. The oxides acted as competing sites for Y segregation, thus dampening the level of enrichment that could be achieved at the grain boundary. The Y atoms expelled from the oxides, having only been displaced a short distance, were inclined to rejoin the oxide due to both the supersaturation of Y in Fe and the preferential transport of Y to the oxides. The disproportionate amount of Ti atom loss in the oxides compared to Y atoms provides further evidence of a Kirkendall effect, where Y atoms enrich in the oxides while Ti is driven to deplete from the oxide region into the Fe matrix. Point defects still transport the Ti atoms to the oxides but at slower rates than Fe and Y, resulting in a net loss of Ti within the composition of the oxide

In visual observations of the oxide under irradiation, there was no oxide precipitation in the region around the grain boundary, and only a few stable precipitates were found elsewhere. Instead, the free solutes tended to rejoin the pre-existing oxides to heal the damage from irradiation. While there was a greater concentration of free solutes in the Fe matrix, there was little increase in solute concentration in the region around the simulated grain boundary. The oxides closer to the grain boundary were even more susceptible to dissolution. This behavior aligns with the theory that the defect population helps stabilize oxides, as the region around the defect sink has a lower average point defect concentration than the rest of the system. Consequently, the oxides in the grain boundary region do not heal as quickly and, therefore, suffer dissolution under irradiation. The binding of vacancies to the oxide interfaces can also play a role in segregation behavior, as it has been noted that similar vacancy binding [51] limits the migration of point defects, making them more susceptible to recombination instead of traveling to the sink (and taking solutes with them) for annihilation. Therefore, there would be less segregation at the grain boundaries.

4.2.6. Replication of the 14YWT neutron experiment.

The oxide response to irradiation in the 14YWT replication simulations showed good agreement with the findings from Aydogan [42]. There was a marginal decrease in the average size of the oxides, and none of the oxides dissolved in the face of irradiation damage. However, occasional short-lived oxides formed from the irradiation debris that quickly dissolved under irradiation. These results display the inherent durability of oxides in the environments required in nuclear reactors.

With the KMC model showing agreement with experimental observations during both heat treatment and irradiation, it opens up possibilities for more extensive research projects, possibly with an artificial intelligence component. The model can also be further extended to observe more phenomena that take place under irradiation. One possibility for this model could be the insertion of transmutation He into the system. Then, observations can be made of the nucleation of He bubbles in the NFAs and the extent to which the nano-oxides inhibit the formation of He bubbles at the grain boundaries.

## 5. Conclusion

A Kinetic Monte Carlo model was created and parameterized for the Fe-Ti-Y-O system. Several simulations of the nucleation and growth stages of nano-oxide $Y_2Ti_2O_7$ formation along the grain boundary were conducted for temperatures of 1023 K, 1123 K, and 1223 K for the 12YWT alloy. The characteristics of the oxides formed were in good agreement with experimental results in size, number density, and shape for the bulk precipitation. The influence of the grain boundary on oxide characteristics was found to be limited in this system. Later, the model subjected the same oxides to neutron irradiation, and the changes in oxide size were recorded. The results found that the oxides were stable against irradiation but suffered a gradual decline in size as the material became more damaged. This stability was true at all temperatures and dose rates studied. An attempt was made to replicate findings for a neutron irradiation experiment of 14YWT and found good agreement with the findings, showing little change in oxide characteristics.

Acknowledgements

This work was supported by the Nuclear Regulatory Commission Fellowship Grant NRC-HQ-84-14-G-0035.

Conflicts of Interest

The authors declare no conflicts of interest with the present work.

Supporting Information:

Appendix A: Neutron Irradiation Mechanism

With the adoption of the 'dpa' unit in the literature, neutron absorption cross-sections and neutron fluxes are not required. It has become common to express the rate of accumulated damage from irradiation as a simple dpa/s. While the damage rate in any material will vary by depth, for small targets like the simulation box, the dpa/s remains practically constant across the entire volume. The probability of a neutron strike, denoted as $\Gamma_{neutStrike}$, is set to the user-input dose rate $G_{dpa}$ in dpa/s and adjusted using the ratio of the number of lattice points in the box $N_{box}$ to the average number of displacements in a cascade $\overline{N_d}$. This ensures that the correct number of neutron strikes occurs to achieve the right dpa over the specified time period.

$$\Gamma_{neutStrike} = G_{dpa} \times \frac{N_{box}}{\overline{N_d}} \quad \text{(A1)}$$

$\overline{N_d}$ from the resulting cascade is estimated by the following equation:

$$\overline{N_d} = \frac{\Lambda E_n}{4 * E_d} \quad \text{(A2)}$$

$$\Lambda = \frac{4A}{(1+A)^2} \quad \text{(A3)}$$

$E_n$ is the initial energy of the neutron, A is the atomic mass number, and $E_d$ is the threshold displacement energy. For Fe, the $E_d$ is set to 40 eV based on literature values. Inside the oxides, the threshold displacement energy for the atoms is set to 57 eV, which is tied to other irradiation studies [52]. It should be noted that these formulas represent an average rather than a set constant [53]. This is a general relationship from Kinchin and Pease [54] and has been found to overestimate the number of displacements from neutron irradiation. Thus, the neutron irradiation is validated by tracking the number of displacements generated by a single displacement cascade and averaging over several trials. For a single 1 MeV neutron striking an all-Fe matrix, the resulting displacement cascade should generate approximately 430 displacements if Eq. A2 and Eq. A3 are applied. A validation procedure was developed and utilized below.

Validation Procedure:

1. Construct a simulation box of pure Fe measuring 100x100x100 lattice sites without any defects.
2. Track 3000 individual displacement cascades caused by 1 MeV neutrons in the box.

3. Find the average $\overline{N_d}$ of displacements per cascade in the KMC and compare it to the predicted value.

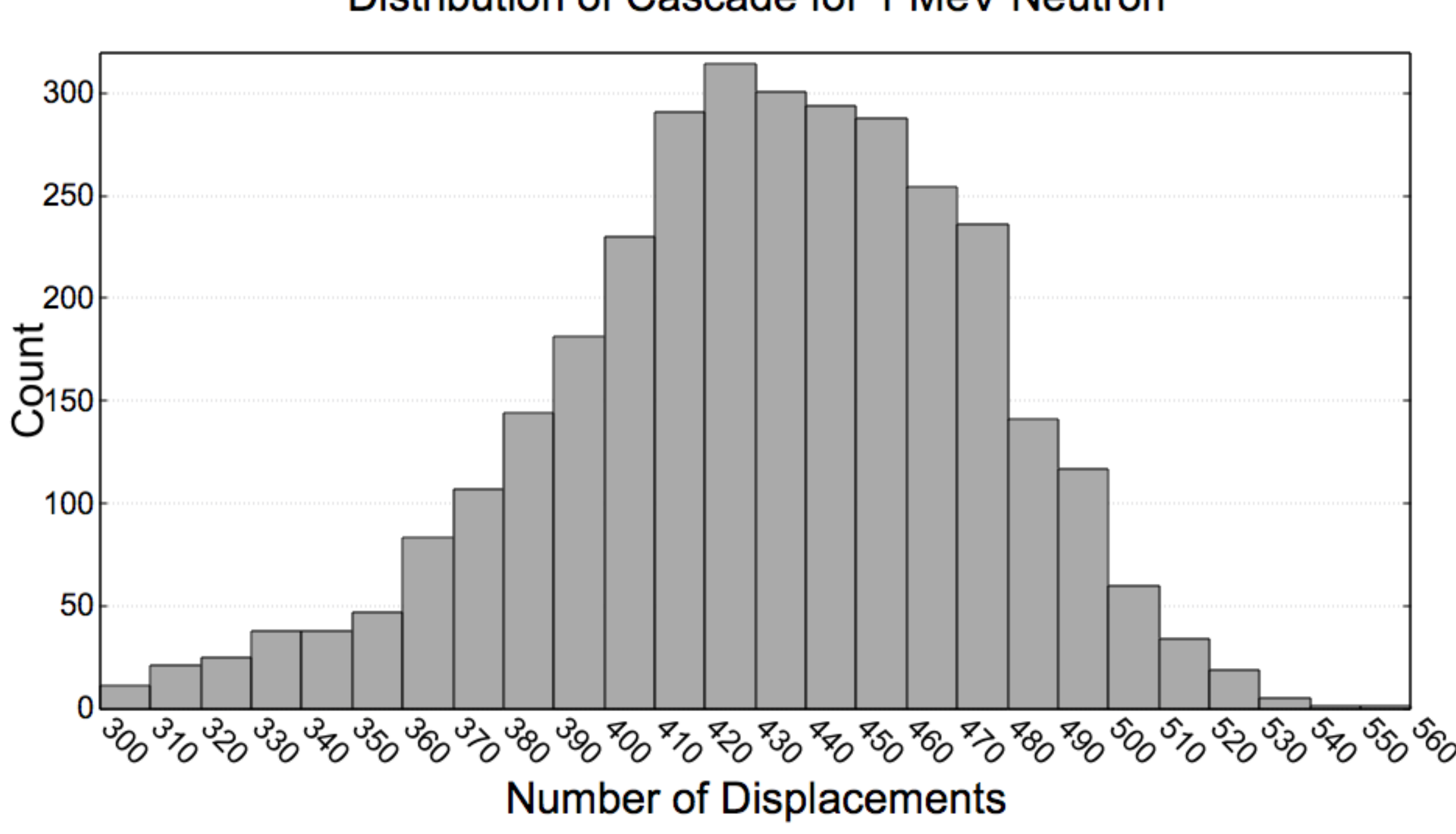

**Figure A1**: Histogram of the number of displacements produced by each 1MeV neutron strike.

Figure A1 displays a histogram of the collected number of displacements in each of the 3000 neutron strikes. The average $\overline{N_d}$ was found to be 430.9 displacements per neutron strike in pure Fe, which aligns with the prediction from the formulas. Therefore, the displacement cascade mechanism is proven to be usable for the KMC.